\documentclass[%
 reprint,
 amsmath,amssymb,
 aps,
 nofootinbib
]{revtex4-1}

\usepackage{graphicx}
\usepackage{xcolor}
\usepackage{dcolumn}
\usepackage{bm}
\usepackage{braket}
\usepackage{amsmath}
\usepackage{amssymb}
\usepackage{enumerate}
\usepackage{booktabs}
\usepackage{txfonts}
\usepackage[update,prepend]{epstopdf}
\usepackage[english]{babel}
\usepackage{listings}
\usepackage{qcircuit}
\usepackage[subrefformat=parens]{subcaption}
\usepackage{caption}
\usepackage{multirow}
\newcommand{\argmin}{\mathop{\rm argmin}\limits}

\makeatletter
\def\hlinewd#1{%
	\noalign{\ifnum0=`}\fi\hrule \@height #1 %
	\futurelet\reserved@a\@xhline}
\makeatother

\begin{document}

\preprint{APS/123-QED}

\title{Quantum Pricing with a Smile: Implementation of Local Volatility Model on Quantum Computer}

\author{Kazuya Kaneko}
\email{kazuya-kaneko@fintec.co.jp}
\affiliation{Mizuho-DL Financial Technology Co., Ltd.\\ 2-4-1 Kojimachi, Chiyoda-ku, Tokyo, 102-0083, Japan}

\author{Koichi Miyamoto}
\email{koichi-miyamoto@fintec.co.jp}
\affiliation{Mizuho-DL Financial Technology Co., Ltd.\\ 2-4-1 Kojimachi, Chiyoda-ku, Tokyo, 102-0083, Japan}

\author{Naoyuki Takeda}
\email{naoyuki-takeda@fintec.co.jp}
\affiliation{Mizuho-DL Financial Technology Co., Ltd.\\ 2-4-1 Kojimachi, Chiyoda-ku, Tokyo, 102-0083, Japan}

\author{Kazuyoshi Yoshino}
\email{kazuyoshi-yoshino@fintec.co.jp}
\affiliation{Mizuho-DL Financial Technology Co., Ltd.\\ 2-4-1 Kojimachi, Chiyoda-ku, Tokyo, 102-0083, Japan}

\date{\today}

\begin{abstract}
Applications of the quantum algorithm for Monte Carlo simulation to pricing of financial derivatives have been discussed in previous papers.
However, up to now, the pricing model discussed in such papers is Black-Scholes model, which is important but simple.
Therefore, it is motivating to consider how to implement more complex models used in practice in financial institutions.
In this paper, we then consider the local volatility (LV) model, in which the volatility of the underlying asset price depends on the price and time.
We present two types of implementation.
One is the {\it register-per-RN} way, which is adopted in most of previous papers.
In this way, each of random numbers (RNs) required to generate a path of the asset price is generated on a separated register, so the required qubit number increases in proportion to the number of RNs.
The other is the {\it PRN-on-a-register} way, which is proposed in the author's previous work.
In this way, a sequence of pseudo-random numbers (PRNs) generated on a register is used to generate paths of the asset price, so the required qubit number is reduced with a trade-off against circuit depth.
We present circuit diagrams for these two implementations in detail and estimate required resources: qubit number and T-count.
\end{abstract}

\pacs{Valid PACS appear here}
                              
\maketitle

\section{\label{sec:intro}Introduction}

With recent advances of quantum computing technologies, researchers are beginning considering how to utilize them in industries.
One major target is finance (see \cite{Orus} for a review).
Since financial institutions are performing enormous tasks of numerical calculation in their daily works, speed-up of such tasks by quantum computer can bring significant benefits to them.
One of such tasks is pricing of financial derivatives\footnote{As textbooks of financial derivatives and pricing of them, we refer to \cite{Hull, Shreve1, Shreve2}}.
Financial derivatives, or simply derivatives, are contracts in which payoffs are determined in reference to the prices of underlying assets at some fixed times.
Large banks typically have a huge number of derivatives written on various types of assets such as stock price, foreign exchange rate, interest rate, commodity and so on.
Therefore, pricing of derivatives is an important issue for them.

In derivative pricing, we represent random movements of underlying asset prices using stochastic processes and calculate a derivative price as a expected value of the sum of payoffs discounted by the risk-free interest rate under some specific probability measure.
In order to calculate the expected value, Monte Carlo simulation is often used.
There are quantum algorithms for Monte Carlo simulation\cite{Montanaro, Suzuki}, which bring quadratic speed-up compared with that on classical computers and there already exists some works which discuss application of such quantum algorithms to derivative pricing\cite{Rebentrost,Stamatopoulos,RamosCalderer}.
However, in order to bring benefits to practice in finance, previous works have some room to be extended.
That is, previous works consider the Black-Scholes (BS) model\cite{BlackScholes,Merton}.
Although the BS model is the pioneering model for derivative pricing and still used in many situations in today's financial firms, it is insufficient to consider only the BS model as an application target of Monte Carlo for practical business for some reasons.
First, for various types of derivatives, market prices of derivatives are inconsistent with the BS model.
This phenomenon is called {\it volatility smile}, which we will explain in Section \ref{sec:LV}.
In order to precisely price derivatives taking into account volatility smiles, financial firms often use models which have more degree of freedom than the BS models.
Second, the BS model is so simple that analytic formulae are available for the price of some types of derivatives in the model.
In such cases, Monte Carlo simulation is not necessary.
Actually, banks use Monte Carlo simulation mainly for complex models which can take into account volatility smiles.
Although it is the natural first step to consider Monte Carlo in the BS model, the above points motivate us to consider how to apply quantum algorithms for Monte Carlo to the advanced models. 

In this paper, we will focus on one of the advanced models, that is, the {\it local volatility (LV)} model\cite{Dupire}.
The LV model, which we will describe later, is the model in which a volatility of an asset price depends on the price itself and time, so the BS model is included in this category as a special case.
With degrees of freedom to adjust the function form of volatility, the LV model can make derivative prices consistent with volatility smiles.
So this model is widely used to price derivatives, especially {\it exotic derivatives}, which have complex transaction terms such as early redemption, in many banks.

In order to price a derivative by Monte Carlo simulation, we generate {\it paths}, that is, random trajectories of time evolution of asset prices, then calculate the expectation value of the sum of discounted payoffs which arise in each path.
Since we cannot generate continuous paths on computers, we usually consider evolutions on a discretized time grid, using a random number (RN) for each time step, which represents the stochastic evolution in the step.
In this paper, we focus on how to implement such a time evolution in the LV model on quantum computers.

We can consider two ways to implement the time evolution.
In this paper, we call them the {\it register-per-RN} way and the {\it PRN-on-a-register} way.
The difference between them is how to generate RNs required to generate a path.
The register-per-RN is adopted in previous papers\cite{Rebentrost,Stamatopoulos,RamosCalderer}.
In this way, following the procedure described in, e.g., \cite{Grover}, one creates a superposition of bit strings which correspond to binary representations of possible values of a RN, where the probability amplitude of a each bit string is the square root of the possibility that the RN take the corresponding value.
The point is that one register is used for one RN, so the required qubit number is proportional to the number of RNs used to generate a path.
This can be problematic in terms of qubit number when many RNs are required.
The number of RNs is equal to that of time steps times and that of underlying assets\footnote{In this paper, we consider arbitrage-free and complete markets, standard assumptions for derivative pricing, so the number of stochastic factors is equal to that of assets. For details, see \cite{Shreve1,Shreve2}.}.
The number of time steps can be large for derivatives with long maturity.
Maturity can be as long as 30 years, so if we take time grid points monthly, the total number of time steps is 360.
The number of underlying assets can be multiple, and furthermore, there are situations where we must simultaneously consider assets concerning different derivative contracts in a portfolio, for example, XVA\footnote{XVA is the term which collectively represents various types of Value Adjustment on derivative prices, for example, credit value adjustment (CVA), price subtraction taking into account the default of the counterparty. In this paper, we ignore such technical issues. If you are interested, see \cite{Gregory}.}.
Assuming that the number of asset is $\mathcal{O}(10)$ and that of time steps is $\mathcal{O}(100)$, the total number of RNs becomes $\mathcal{O}(10^3-10^4)$.
If we use a register with $\mathcal{O}(10)$ qubits for each RN, the total qubit number can be $\mathcal{O}(10^5)$ easily.
The current state-of-art quantum computers have only $\mathcal{O}(10)$ qubits\cite{Arute}.
Even if we obtain large-scale fault-tolerant machines in the future, the large qubit overhead might be required to make a logical bit (see \cite{Campbell} as a review and references therein).
Therefore, calculations which require the large number of qubits as above might be prohibitive even in the future.
This situation is similar to credit portfolio risk measurement, which is described in \cite{Miyamoto}.

We are then motivated to consider the PRN-on-a-register way, which is proposed in \cite{Miyamoto}.
In this way, one does not create RNs on different register, but generates a sequence of pseudo-random number (PRN) on a register.
At each time step, the PRN sequence is progressed and its value is used to evolve the asset price.
Therefore, the required qubit number does not depend on the number of RNs and is largely reduced.
The drawback is the circuit depth.
Here, we define the circuit depth as the number of layers consisting of gates on distinct qubits that can be performed simultaneously, as T-depth used in \cite{Amy2,Selinger}.
Since calculations to update the PRN is sequentially performed on a register, the circuit depth is now proportional to the number of RNs.
Since in the fault-tolerant computation some kinds of gates, for example T-gates in the Clifford+T gate set, can take long time to be run\cite{Bravyi,Fowler}, the sequential run of such gates might be also prohibitive in terms of calculation time\cite{Egger}.
At any rate, in the current stage, where it is difficult to foresee the spec of future quantum computers, we believe that it is meaningful to consider the implementation which saves qubits but consumes depth as a limit.

When it comes to the LV model, the PRN-on-a-register way becomes more motivating, since its disadvantage on the circuit depth compared with the register-per-RN way is alleviated.
In the LV model, the volatility varies over time steps depending on the asset price, so the calculation for the time evolution is necessarily stepwise\footnote{In the multi-asset case, parallel computing over assets is possible in the register-per-RN way.}.
Therefore, the PRN-on-a-register way and the register-per-RN way are equivalent with respect to this point, that is, the circuit depth is proportional to the time step number in both ways.
This is different from the situation in credit portfolio risk management \cite{Egger}, where, in the register-per-RN way, a register is assigned to each random number which determine whether each obligor defaults or not and parallel processing on different registers reduces circuit depth.

In this paper, we design the quantum circuits in the above two way in the level of elementary arithmetic.
In doing so, we follow the policies of the two ways to the extent possible.
That is, not only with respect to RNs but also in other aspects, we try to reduce qubits accepting some additional procedures in the PRN-on-a-register way, and vice versa in the register-per-RN way.
For example, in the PRN-on-a-register way, we have to intermediately output the  information of the volatility used to evolve the asset price at each time step and clear it by the next step.
Otherwise, we need a register to hold the information per step and the required qubit number becomes proportional to the number of time steps.
It is nontrivial to implement such a procedure and we will present how to do this later.
Note that such clearing procedure is unnecessary in the register-per-RN way.

We then estimate the resources to implement the proposed circuits.
We focus on two metrics: qubit number and T-count.
As mentioned above, we see that the qubit number in the PRN-on-a-register way is independent from the time step number and much less than the register-per-RN way.
The T-count is proportional to the time step number in the both ways.
We see that in some specific setting the both ways yield the T-counts of same order of magnitude, except that in the PRN-on-a-register way is larger by some $\mathcal{O}(1)$ factor.

The rest of this paper is organized as follows.
Section \ref{sec:LV} and \ref{sec:QMC} are preliminary sections, the former and the latter briefly explain the LV model and the quantum algorithm for Monte Carlo simulation, respectively.
In section \ref{sec:circuit}, we present the circuit diagram in the two way.
In section \ref{sec:resources}, we estimate qubit number and T-count of the proposed circuits.
Section \ref{sec:con} gives a summary.

\section{\label{sec:LV}LV model}

In this paper, we consider only the single-asset case, since it is straightforward to extend the discussion in this paper to the multi-asset case.

\subsection{pricing of derivatives}
Consider a party A involved in a derivative contract written on some asset.
We let $S_t$ denote a stochastic process which represents the asset price at time $t$, which is set as $t=0$ at the present.
We assume that the payoffs arise at the multiple times $t^{\rm pay}_{i},i=1,2,...$ and the $i$-th payoff is given by $f^{\rm pay}_{i}\left(S_{t^{\rm pay}_{i}}\right)$, where $f^{\rm pay}_i$ is some function which maps $\mathbb{R}$ to $\mathbb{R}$.
The positive payoff means that A receives a money from the counterparty and the negative one means vice versa.
For example, the case where A buys an European call option with the strike $K$ corresponds to
\begin{equation}
f^{\rm pay}_{1}(S_{t^{\rm pay}_{1}})=\max\left\{S_{t^{\rm pay}_{1}}-K,0\right\} \label{eq:PayECall}
\end{equation}
with a single payment date $t^{\rm pay}_{1}$. 
Note that this type of derivative contract is too simple to cover all trades in financial markets.
For example, {\it callable} contracts, in which either of the parties has a right to terminate the contract at some times, are widely dealt in markets.
We leave studies for such {\it exotic} derivatives for future works and, in this paper, concider only those which can be expressed in the above form.

Following the theory of arbitrage-free pricing, the price $V$ of the contract for A is given as follows \cite{Shreve1,Shreve2}:
\begin{equation}
V=E\left[\sum_i f^{\rm pay}_{i}\left(S_{t^{\rm pay}_{i}}\right)\right], \label{eq:price}
\end{equation}
where $E[\cdot]$ represents the expectation value under some probability measure, the so-called {\it risk-neutral measure}.
Here and hereafter, we assume that the risk-free interest rate is 0 for simplicity.

\subsection{LV model}

In the LV model, the evolution of the asset price is modeled by the following stochastic differential equation (SDE)
\begin{equation}
dS_t = \sigma(t,S_t)dW_t \label{eq:SDE}
\end{equation}
in the risk-neutral measure\footnote{Note that the drift term does not exist since we are now assuming the risk-free rate is 0.}
$W_t$ is the Wiener process which drives $S_t$.
$dX_t$ is the increment of a stochastic process $X_t$ over an infinitesimal time interval $dt$.
The deterministic function $\sigma:[0,\infty)\otimes \mathbb{R} \rightarrow [0,\infty)$ is the local volatility.
Note that the BS model corresponds to the case where
\begin{equation}
\sigma(t,S) = \sigma_{\rm BS}S, \label{eq:LVinBS}
\end{equation}
where $\sigma_{\rm BS}$ is a positive constant, which we hereafter call a {\it BS volatility}.

The LV model was proposed to explain {\it volatility smile}.
In order to describe this, let us define {\it implied volatility} first.
In the BS model, a price of a European call option with strike $K$ and maturity $T$ at $t=0$ is given by the following formula:
\begin{eqnarray}
V_{\rm call,BS}(T,K,S_0,\sigma_{\rm BS}) & = & \Phi_{\rm SN}(d_1)S_0 - \Phi_{\rm SN}(d_2)K \nonumber \\
d_1 & = & \frac{1}{\sigma_{\rm BS}\sqrt{T}}\left[\ln\left(\frac{S_0}{K}\right)+\frac{1}{2}\sigma_{\rm BS}^2T\right] \nonumber \\
d_2 & = & d_1 - \sigma_{\rm BS}\sqrt{T},
\end{eqnarray}
where $\Phi_{\rm SN}$ is the cumulative distribution function of the standard normal distribution.
We can price the option if we determine the BS volatility.
Conversely, given the market price of the option $V_{\rm call,mkt}(T,K)$, we can reversely calculate the BS volatility.
That is, we can define the following function of $K$ and $T$:
\begin{eqnarray}
\sigma_{\rm IV} & : & (T,K) \mapsto \nonumber \\
& & \sigma_{\rm IV}(T,K) \ s.t. \ V_{\rm call,BS}(T,K,S_0, \sigma_{\rm IV}(T,K)) = V_{\rm call,mkt}(T,K). \nonumber \\
& & \label{eq:IV}
\end{eqnarray}
We call BS volatilities drawn back from the market option prices by (\ref{eq:IV}) as implied volatilities.

If the market is described well by the BS model, implied volatilities $\sigma_{\rm IV}(T,K)$ take a same value for any $K$ and $T$.
Although this is the case for some markets, $\sigma_{\rm IV}(T,K)$ varies depending on $K$ and $T$ in many markets.
Especially, if $\sigma_{\rm IV}(T,K)$ depends on $K$, it is said that we observe the volatility smile for the market.

Volatility smiles mean that possible scenarios of asset price evolution in the BS model do not match those which market participants consider.
For example, if market participants think that extreme scenarios, big crashes or sharp rises, are more possible than the BS model predicts, the volatility smile arises.
In fact, it is often said that the Black Monday, the big crash in the stock markets at 1987, was one of triggers of appearance of volatility smiles.

With the LV model, we can make European option prices given by the model consistent with any market prices, as long as there is no arbitrage in the market.
This is intuitively apparent since we can expect that the degree of freedom of the local volatility $\sigma(t,S)$ as a two-dimensional function is available to reproduce the two-dimensional function $V_{\rm call,mkt}(T,K)$.
In fact, if we can get $V_{\rm call,mkt}(T,K)$ as a function, that is, the market option prices for continuously infinite strikes and maturities, we can determine $\sigma(T,K)$ which reproduces $V_{\rm call,mkt}(T,K)$ as follows\cite{Dupire}:
\begin{equation}
\sigma^2(T,K) = 2\frac{\frac{\partial}{\partial T} V_{\rm call,mkt}(T,K)}{\frac{\partial^2}{\partial K^2} V_{\rm call,mkt}(T,K)}.
\end{equation} 

In reality, the market option prices are available only for several strikes and maturities.
Therefore, in the practical business, we usually use a specific functional form of $\sigma(t,S)$ with degrees of freedom sufficient to reproduce several available market option prices.
In this paper, we use the following form.
First, we set the $n_t$ grid points in the time axis, $t_0=0<t_1<t_2<...<t_{n_t}$.
Second, we set the $n_S$ grid points in the asset price axis for each time grid point, that is, $s_{i,1},...,s_{i,n_S}$ for $t_i$.
Then, $\sigma(t,S)$ is set as follows:
\begin{equation}
\sigma(t,S) = a_{i,j}S + b_{i,j} \ ; {\rm for} \ s_{i,j-1} \leq S < s_{i,j}, j=1,...,n_S+1 \label{eq:LV}
\end{equation}
for $t_{i-1} \leq t < t_i$, where $a_{i,j},b_{i,j}$ are constants satisfying $\sigma(t,S)>0$ for any $t$ and $S$ and $s_{i,0}=-\infty,s_{i,n_S+1}=+\infty$.
In other words, the two-dimensional space of $(t,S)$ is divided in the direction of $t$ and in each region $\sigma(t,S)$ is set to a function which is piecewise-linear with respect to $S$.
In this paper, we assume that $a_{i,j},b_{i,j}$ are preset to the value for which the option prices in the LV model, which we here write as $V_{\rm call,LV}(T,K,S_0,\{a_{i,j}\},\{b_{i,j}\})$, match the market prices by some standard.
For example, they can be set to
\begin{equation}
\argmin_{\{a_{i,j}\},\{b_{i,j}\}} \sum_I{\left[V_{\rm call,LV}(T_I,K_I,S_0,\{a_{i,j}\},\{b_{i,j}\}) - V_{\rm call,mkt}(T_I,K_I)\right]^2},
\end{equation} 
where $(T_I,K_I)$'s are several sets of maturity and strike for which the market option prices are available.

\subsection{Monte Carlo simulation in the LV model}

We here describe how to calculate the derivative price (\ref{eq:price}) in the LV model by Monte Carlo simulation.

First, we have to discretize the time into sufficiently small meshes, since we can deal with the continuous time on neither classical nor quantum computers.
For simplicity, we set the time grid points to the above $t_i$'s, those for the LV function.
Then, the time evolution given by (\ref{eq:SDE}) is approximated as follows:
\begin{equation}
\Delta S_{t_i} := S_{t_{i+1}} - S_{t_i} \approx \sigma(t_i,S_{t_i})\sqrt{\Delta t_i}w_i, \Delta t_i=t_{i+1}-t_i, \label{eq:EM}
\end{equation}
where $w_1,...,w_{n_t}$ are independent standard normal random numbers (SNRNs).
Among various ways to discretize the SDE, we here adopt the Euler-Maruyama method \cite{Maruyama}.

Even after time discretization, we cannot consider all of continuously infinite patterns of SNRNs.
One solution for this is discretized approximation of SNRNs.
We can choose the finite numbers of the grid points and assign probability to each point so that standard normal distribution is approximately reproduced.
Now, the patterns of discretized SNRNs are finite, so we can approximate (\ref{eq:price}) as
\begin{equation}
V\approx \sum_{n} p_n  \sum_i f^{\rm pay}_{i}\left(S^{(n)}_{t^{\rm pay}_{i}}\right),  \label{eq:priceMCRN}
\end{equation}
where $p_n$ is the probability that the $n$-th pattern of values of SNRNs are realized and $S^{(n)}_t$ is the asset price at time $t$ in the $n$-th pattern.

There are some possible ways to take petterns considered in (\ref{eq:priceMCRN}).
In the register-per-RN way, we take {\it all} patterns.
If we take $N$ grids to discretize each of $n_t$ SNRNs, the number of possible patterns of SNRNs is $N^{n_t}$.
Although this is exponentially large, quantum computers can take into account all patterns with a polynomial number of qubits by quantum superposition.

On the other hand, this cannot be adopted on classical computers, since the number of the SNRN patterns are exponentially large.
Usually, Monte Carlo pricing on classical computers is done in the following way, which the PRN-on-a-register way is also based on.
We consider {\it sampled} patterns of SNRNs.
That is, we generate finite but sufficiently many sample sets of $(w_1,...,w_{n_t})$ and use them to generate sample paths of the asset price which evolves according to (\ref{eq:EM}).
We then approximate (\ref{eq:price}) by the average of sums of payoffs in sample paths,
\begin{equation}
V\approx \frac{1}{N_{\rm path}}\sum_{n=1}^{N_{\rm path}} \sum_i f^{\rm pay}_{i}\left(S^{(n)}_{t^{\rm pay}_{i}}\right), \label{eq:priceMC}
\end{equation}
where $S^{(n)}_t$ is the value of the asset price at time $t$ on the $n$-th sample path and $N_{\rm path}$ is the number of sample paths.

\section{\label{sec:QMC}Quantum Algorithm for Monte Carlo Simulation}

\subsection{outline of the algorithm}

We here review the quantum algorithm for Monte Carlo simulation\cite{Montanaro,Suzuki}.
It can be divided into the following steps.
First, we create a superposition of possible values of a random number used to calculate a sample value of the integrand on a register.
If multiple random numbers are necessary to calculate the integrand, one register is assigned per random number.
As mentioned above, continuous random numbers must be approximated in some discretized way.
Second, we calculate the integrand into another register, using the random numbers.
Note that the results for many patterns of random numbers are simultaneously calculated in quantum parallelism.
Third, by controlled rotation, the integrand value is reflected into the amplitude of the ancilla.
Finally, amplitude estimation \cite{Bassard,Suzuki,Nakaji} on the ancilla gives the expectation value of the integrand.

The quantum state is transformed as follows:
\begin{eqnarray}
& & \ket{0}\ket{0}\ket{0} \nonumber \\
& \rightarrow & \left(\sum_i{\sqrt{p_i}\ket{x_i}}\right)\ket{0}\ket{0} \nonumber \\
& \rightarrow & \left(\sum_i{\sqrt{p_i}\ket{x_i}}\ket{f(x_i)}\right)\ket{0} \nonumber \\
& \rightarrow & \sum_i{\sqrt{p_i}\ket{x_i}}\ket{f(x_i)}\left(\sqrt{1-f(x_i)}\ket{0}+\sqrt{f(x_i)}\ket{1}\right)  \nonumber \\
& &. \label{eq:QMC}
\end{eqnarray}
Here, the first, second and third kets correspond to the random number registers, the integrand register and the ancilla, respectively.
$x_i$ represents the binary representation of values of random numbers in the $i$-th pattern and $p_i$ is the probability that it realizes.
$f$ is the integrand and $f(x_i)$ is its value for $x_i$. 
Note that the probability to observe 1 on the ancilla is $\sum_i{p_if(x_i)}$, the integral value which we want.
Although we do not explain how to estimate the probability amplitude in this paper, it is studied in many papers.
For example, see \cite{Bassard,Suzuki,Nakaji}.
Using such methods, we can estimate the integral with the statistical error which decays as $\mathcal{O}(N^{-1})$, where $N$ is the number of oracle calls. 
This decay rate is quadratically faster than that in the classical algorithm, $\mathcal{O}(N^{-1/2})$.

\subsection{the scheme using the PRN generator}
We here briefly review the quantum way for Monte Calro simulation using the PRN generator.
The calculation flow for the current problem, the time evolution of asset price in the LV model, based on this way is described in Section \ref{sec:PRNway}.

It is proposed in \cite{Miyamoto} in order to reduce the required qubits to generate RNs in the application of the quantum algorithm for Monte Carlo to extremely high-dimensional integrations.
When it is neccesary to generate many RNs to compute the integrand, the naive way, in which we assign a register to each RN and create a superposition of possible values, leads to the increase of qubit numbers in proportion to the number of RNs.
In order to aviod this, we can adopt the following way.
First, we prepare two registers, $R_{\rm samp}$ and $R_{\rm PRN}$.
Then, we create a superposition of integers, which specify the start point of the PRN sequence, on $R_{\rm samp}$.
With the start point, we sequentially generate PRNs on $R_{\rm PRN}$.
This is possible because a PRN sequence is a deterministic sequence whose recursion equation is explicitely given, and in \cite{Miyamoto} we gave the implementation of one of PRN generators on quantum circuits.
Using the PRNs, we compute the integrand step by step, which corresponds to time evolution of the asset price and calculation of payoffs in this paper.
Finally, the expectation value of the integrand is estimated by quantum amplitude estimation.
In this way, since we need only $R_{\rm samp}$ and $R_{\rm PRN}$ to generate PRNs, the required qubit number is now independent from the number of RNs and much smaller than the naive way.
The drawback is the increase of the circuit depth.

\section{\label{sec:circuit}Circuit Design}

Now, we present quantum circuits for time evolution of an asset price in the LV model in the two ways: PRN-on-a-register and register-per-RN.

\subsection{\label{sec:elemgate}elementary gate}

Before we present circuits, we here list up elementary gates we use.

\begin{itemize}
	\item Adder: $\ket{x}\ket{y}\rightarrow \ket{x+y}\ket{y}$
	\item Controlled Adder:\\ $\ket{c}\ket{x}\ket{y}\rightarrow \begin{cases}
	\ket{c}\ket{x+y}\ket{y} \ ; \ {\rm for} \ c=1 \\ \ket{c}\ket{x}\ket{y} \ ; \ {\rm for} \ c=0
	\end{cases} $
	\item Multiplier: $\ket{x}\ket{y}\ket{z}\rightarrow \ket{x}\ket{y}\ket{z+xy}$
	\item Divider: $\ket{x}\ket{y}\ket{0}\rightarrow \ket{x}\ket{y}\ket{x/y}$
\end{itemize}
We here simply assume their existence.
Actually, implementation of such elementary arithmetics are widely studied in previous works: see, for example, \cite{Vedral,Beckman,Draper,Cuccaro,Takahashi,VanMeter,Draper2,Takahashi2,Portugal,AlvarezSanchez,Takahashi3,Thapliyal,Thapliyal2,Lin,Babu,Jayashree,MunozCoreas,Khosropour,Jamal,Dibbo,Thapliyal3}.
We will discuss the implementation in Section \ref{sec:elemgateres}.

With these, we can construct other types of arithmetic we use.
For example, subtraction $\ket{x}\ket{y}\rightarrow \ket{x-y}\ket{y}$ can be done as addition by the method of 2's complement.
Comparison $\ket{x}\ket{y}\ket{z}\rightarrow \ket{x}\ket{y}\ket{z\oplus(x>y)}$ can be done as subtraction in 2's complement method, since the most significant bit represents whether the result of subtraction is positive or negative.
So a comparator is constructed as two adder, including uncomputation.

Note that the above multiplier uses two registers as operands and outputs the product into another register.
Most of previously proposed multipliers are of this output-to-other type.
On the other hand, we also need the self-update type of multiplier which updates either of input registers with the product, otherwise we have to add a register for each multiplication and qubit number explodes.
Such a operation is realized by the following trick.
When we want to multiply $x$ by $y$, given the two registers holding $x$ and $y$ and an ancilla register, we can do:
\begin{equation}
\ket{x}\ket{y}\ket{0} \rightarrow \ket{x}\ket{y}\ket{xy} \rightarrow \ket{xy}\ket{y}\ket{x} \rightarrow \ket{xy}\ket{y}\ket{0}. \label{eq:xy}
\end{equation}
Here, the first step is output-to-other multiplication.
The second step is swap between the first and third registers, which is not necessary if we change our recognition on which of two register is ancillary at every multiplication.
The third step is the inverse operation of division.

\subsection{the PRN-on-a-register way\label{sec:PRNway}}

\subsubsection{calculation flow}

We first present the calculation flow in the PRN-on-a-register way.
We consider the flow until calculation of the sum of payoffs, which corresponds to from the first to the third line in (\ref{eq:QMC}), since the controlled rotation in the fourth line does not depend on the problem.

In the PRN-on-a-register way, PRNs are used for evolution of the asset price (\ref{eq:EM}).
More concretely, we preselect some sequence of pseudo standard normal random numbers (PSNRNs) and divide it into subsequences, then evolve the asset price using them.

Before we present the calculation flow, we explain some setups.
We prepare the following register:
\begin{itemize}
	\item $R_{\rm samp}$
	
	This is a register where a superposition of integers which determine the start point of the PSNRN sequence.
	We write its qubit number as $n_{\rm samp}$.
	$N_{\rm samp}=2^{n_{\rm samp}}$ is the number of sample paths we generate.
	
	\item $R_W$
	
	This is a register where we sequentially generate PSNRNs.	
	
	\item $R_S$
	
	This is a register where the value of the asset price is stored and which we update for each time step of the evolution, using $R_W$.
	
	\item $R_{\rm payoff}$
	
	This is a register into which the payoffs determined by $R_S$ are added.
	
\end{itemize}
Note that we need some ancillary registers in addition to the above registers.
We assume that the required calculation precision is $n_{\rm dig}$-bit accuracy and  $R_W, R_S, R_{\rm payoff}$ and ancillary registers necessary to update them have $n_{\rm dig}$ qubits.

We assume that the following gates are available to generate a sequence of PSNRNs.
\begin{itemize}
	\item $P_{W}$
	
	This progresses a PSNRN sequence by one step. In other words, it acts on $R_W$ and updates $x_i$ to $x_{i+1}$, where $x_i$ is the $i$-th element of the sequence: $\ket{x_i} \rightarrow \ket{x_{i+1}}$.
	
	\item $J_{W}$
	
	This lets the PSNRN sequence jump to the starting point.
	That is, it is input an integer $i$ on a register and outputs $x_{in_t+1}$ into another register which is initially set to $\ket{0}$: $\ket{i}\ket{0} \rightarrow \ket{i}\ket{x_{in_t+1}}$.
	Remember that $n_t$, the number of time steps, is equal to the number of RNs used to generate one sample path.
	
\end{itemize}
The concrete implementation of these gates are discussed later.

Then, the calculation flow is as follows:
\begin{enumerate}
	\item Initialize all registers to $\ket{0}$ except $R_S$, which is initialized to $\ket{S_{t_0}}$.
	\item Generate a equiprobable superposition of $\ket{0}, \ket{1},...,\ket{N_{\rm{samp}}-1}$, that is, $\frac{1}{\sqrt{N_{\rm{samp}}}}\sum_{i=0}^{N_{\rm{samp}}-1}{\ket{i}_{n_{\rm{PRN}}}}$ on $R_{\rm samp}$. This is done by operating a Hadamard gate to each of $n_{\rm{samp}}$ qubits.
	\item Operate $J_W$ to set $x_{in_t+1}$ to $R_W$, where $i$ is determined by the state of $R_{\rm{samp}}$. These are the starting points of subsequences.
	\item Perform the time evolution (\ref{eq:EM}) using the value on $R_W$. $R_S$ is updated from $\ket{S_{t_0}}$ to $\ket{S_{t_1}}$.
	\item Calculate the payoff at time $t_1$ and add into $R_{\rm payoff}$.
	\item Operate $P_{\rm{PRN}}$ to update $R_{W}$ from $x_{in_t+1}$ to $x_{in_t+2}$.
	\item Iterate operations similar to 4-6 for each time steps until the time reaches $t_{n_t}$.
	\item Finally we obtain a superposition of states in which the value on $R_{\rm payoff}$ is the sum of payoffs in each sample path. Estimate the expectation value of $R_{\rm payoff}$ by methods like \cite{Bassard,Suzuki}. This is an estimate for (\ref{eq:priceMC}).
\end{enumerate}
Here and hereafter, we assume that a payoff arises at each time step, for simplicity.

The flow of state transformation is as follows:
\begin{eqnarray}
&   \ket{0}\ket{0}\ket{S_{t_0}}\ket{0} & \nonumber \\ 
\xrightarrow{2} & \frac{1}{\sqrt{N_{\rm{samp}}}}\sum_{i=0}^{N_{\rm{samp}}-1} & {\ket{i}}\ket{0}\ket{S_{t_0}}\ket{0} \nonumber \\
\xrightarrow{3} & \frac{1}{\sqrt{N_{\rm{samp}}}}\sum_{i=0}^{N_{\rm{samp}}-1} & {\ket{i}} \ket{x_{in_t+1}}\ket{S_{t_0}}\ket{0} \nonumber \\
\xrightarrow{4} & \frac{1}{\sqrt{N_{\rm{samp}}}}\sum_{i=0}^{N_{\rm{samp}}-1} & {\ket{i}}\ket{x_{in_t+1}}\ket{S^{(i)}_{t_1}}\ket{0} \nonumber \\
\xrightarrow{5} & \frac{1}{\sqrt{N_{\rm{samp}}}}\sum_{i=0}^{N_{\rm{samp}}-1} & {\ket{i}} \ket{x_{in_t+1}}\ket{S^{(i)}_{t_1}}\ket{\sum_{j=1}^{1}{f^{\rm pay}_{j}(S^{(i)}_{t_j})}} \nonumber \\
\xrightarrow{6} & \frac{1}{\sqrt{N_{\rm{samp}}}}\sum_{i=0}^{N_{\rm{samp}}-1} & {\ket{i}}\ket{x_{in_t+2}}\ket{S^{(i)}_{t_1}}\ket{\sum_{j=1}^{1}{f^{\rm pay}_{j}(S^{(i)}_{t_j})}} \nonumber \\
\xrightarrow{7} & ... \qquad\quad\quad\quad & \nonumber \\
\xrightarrow{7} & \frac{1}{\sqrt{N_{\rm{\rm{samp}}}}}\sum_{i=0}^{N_{\rm{\rm{samp}}}-1} & {\ket{i}}\ket{x_{in_t+n_t}}\ket{S^{(i)}_{t_{n_t}}}\ket{\sum_{j=1}^{n_t}{f^{\rm pay}_{j}(S^{(i)}_{t_j})}}, \label{eq:calcflowPRN}
\end{eqnarray}
where the first, second, third and fourth kets correspond to $R_{\rm samp},R_{W},R_S$ and $R_{\rm payoff}$, respectively.

\subsubsection{overview of the circuit}

\begin{figure*}[t]
	\begin{center}
		\includegraphics[width=0.7\textwidth]{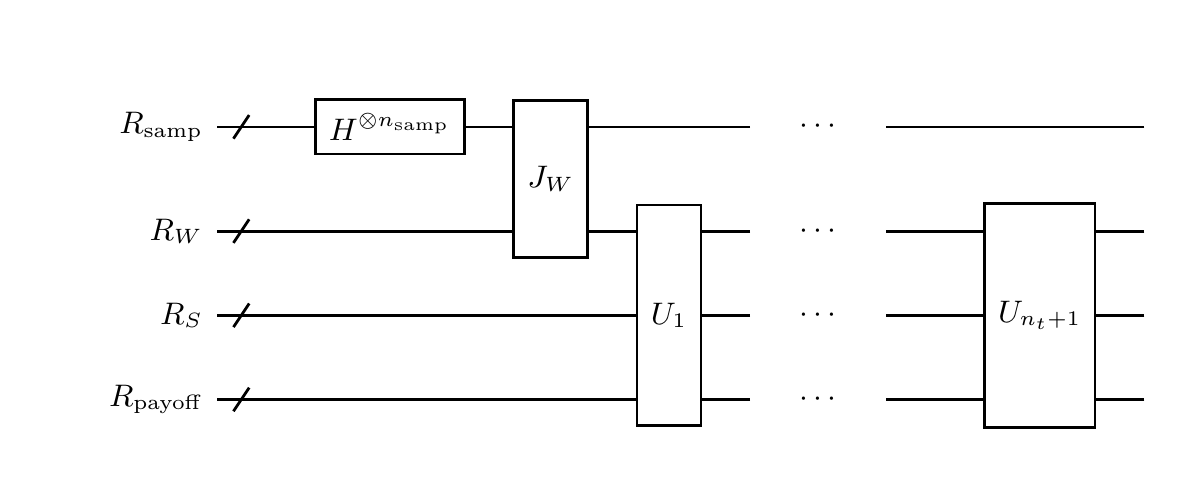}
	\end{center}
	\caption{The overview of the circuit for asset price evolution in the LV model in the PRN-on-a-register way. Here and hereafter, ancillary qubits are sometimes omitted for simple display.}
	\label{fig:overviewPRN}.
\end{figure*}

Schematically, the circuit which realizes the flow (\ref{eq:calcflowPRN}) is as shown in Figure \ref{fig:overviewPRN}.
In the figure, the gate $U_j$ corresponds to the $j$-th step of asset price evolution, that is, the $j$-th iteration of step 4-6 in the above calculation flow.

\begin{figure*}[t]
	\begin{center}
		\includegraphics{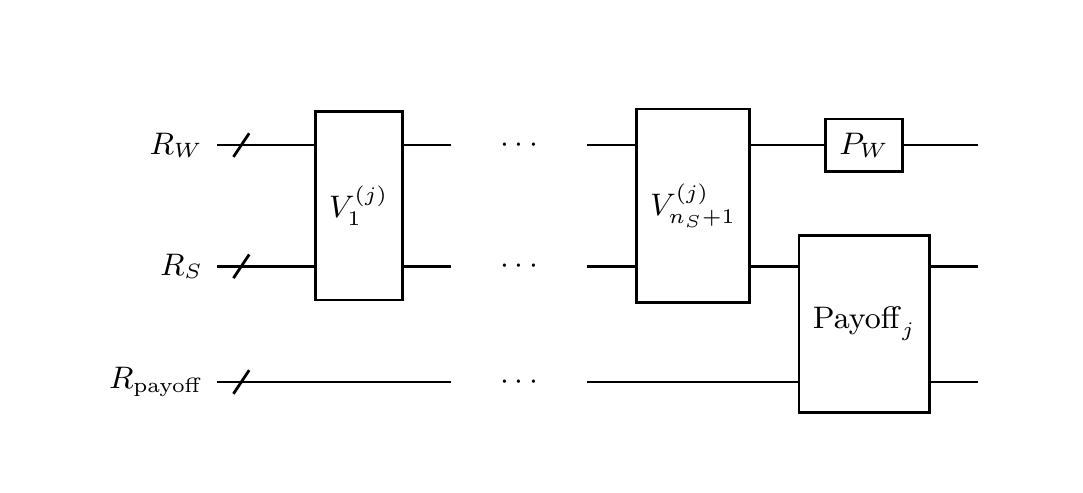}
	\end{center}
	\caption{The overview of the $U_j$, which performs the $j$-th step of asset price evolution, in the PRN-on-a-register way.}
	\label{fig:UtPRN}
\end{figure*}

$U_j$ is implemented as shown in Figure \ref{fig:UtPRN}.
$P_W$ is already explained and the gate ${\rm Payoff}_j$ calculates $f^{\rm pay}_{j}(S^{(i)}_{t_j})$ using $R_S$ and adds it into $R_{\rm payoff}$.
In addition to these, $U_j$ has gates $V^{(j)}_{1},...,V^{(j)}_{n_S}$, which update $R_S$.

The detail of $V^{(j)}_{k}$ is shown in Figure \ref{fig:Vjk}.
This gate (i) checks whether the asset price is in the $k$-th interval $[s_{j,k-1},s_{j,k})$, (ii) if so, update $R_S$ using the LV in the interval, (iii) clears the intermediate output.
It requires ancillary registers $R_{\rm count}, R_{S^\prime}$ and $R_g$.
They have $\lceil\log_2{n_t}\rceil, n_{\rm dig}$ and 1 qubits respectively.
At the start of $V^{(j)}_{k}$, $R_{\rm count}$ takes $\ket{j}$ or $\ket{j+1}$ and the others take $\ket{0}$.
Then the detailed calculation flow is:
\begin{enumerate}
	\item If $R_{\rm count}$ is $j$ and $R_S$ is in $[s_{j,k-1},s_{j,k})$, flip $R_g$.
	\item If $R_g$ is 1, update $R_S$ as 
	\begin{equation}
	S_{t_j}\rightarrow S_{t_{j+1}}=S_{t_j}+(a_{j,k}S_{t_j}+b_{j,k})\sqrt{\Delta t_j}x_{in_t+j} \label{eq:SUp}
	\end{equation}
	using the value $x_{in_t+j}$ on $R_W$ and add 1 to $R_{\rm count}$.
	\item Calculate
	\begin{equation}
	\frac{S_{t_{j+1}}-b_{j,k}\sqrt{\Delta t_j}x_{in_t+j}}{1+a_{j,k}\sqrt{\Delta t_j}x_{in_t+j}}
	\end{equation}
	into $R_{S^\prime}$, using the value on $R_S$ as $S_{t_{j+1}}$ and that on $R_W$ as $x_{in_t+j}$.
	\item If $R_{\rm count}$ is $j+1$ and $R_{S^\prime}$ is in $[s_{j,k-1},s_{j,k})$, flip $R_g$. This uncomputes $R_g$.
	\item Do the inverse operation of 3.
\end{enumerate}

Let us explain the meaning of this flow.
First, $R_{\rm count}$ is necessary as an indicator of whether the $j$-th step of evolution has been already done or not.
Without this, it is possible that the asset price is doubly updated in a time step.
If and only if the $j$-th step has not been done, that is, $R_{\rm count}$ is $j$ and the asset price is in $[s_{j,k-1},s_{j,k})$, the update of the asset price with the LV function $a_{j,k}S+b_{j,k}$ is done.
To do this conditional update, the check result is intermediately output to $R_g$ and the gate corresponding (\ref{eq:SUp}) is operated on $R_S$ under control by $R_g$.
Besides, the increment of $R_{\rm count}$ controlled by $R_g$ is also done, so that $R_{\rm count}$ indicates completion of the $j$-th step if so.
Steps 3-5 is necessary to clear $R_g$.
If the asset price has been updated in Step 2, Step 3 draws back it to the value before the update.
Conversely, we can determine whether the update has been done in Step 2 from the result of Step 3.
That is, for the reason mentioned soon later, the condition that $R_{\rm count}$ is $j+1$ and $R_{S^\prime}$ is in $[s_{j,k-1},s_{j,k})$ after Step 3 is equivalent to the condition that $R_{\rm count}$ is $j$ and $R_S$ is in $[s_{j,k-1},s_{j,k})$ before Step 2.
Therefore, Step 4 flip $R_g$ if and only if it is $\ket{1}$, so it goes back to $\ket{0}$.
In summary, through the sequential operation of $V^{(j)}_1,...,V^{(j)}_{n_S+1}$, $R_S$ is updated only once at the appropriate $V^{(j)}_k$, $R_{\rm count}$ is updated from $\ket{j}$ to $\ket{j+1}$ and all intermediate outputs on ancillary registers are cleared. 

We here mention a restriction on the LV model so that it can be implemented in the PRN-on-a-register way.
Note that through $V^{(j)}_1,...,V^{(j)}_{n_S+1}$, the state is transformed from $\ket{j}\ket{S^{(i)}_{t_j}}$ to $\ket{j+1}\ket{S^{(i)}_{t_{j+1}}}$, where the first and second kets correspond to $R_{\rm count}$ and $R_S$ respectively and other registers are omitted since they are unchanged.
This means that the map from $S^{(i)}_{t_j}$ to $S^{(i)}_{t_{j+1}}$ must be one-to-one correspondence, since unitarity is violated if not.
Actually, this is not so strong restriction.
As shown in Appendix, if we set $a_{i,j},b_{i,j}$ so that $\sigma(t,S)$ is continuous with respect to $S$ and we set $\Delta t_{j}$ is small enough that the increment $\Delta S_{t_j}$ is much smaller than $S_{t_j}$ itself, the above condition is satisfied.

This one-to-one correspondence lets Step 3 work.
That is, since the map between $S^{(i)}_{t_j}$ and $S^{(i)}_{t_{j+1}}$ is one-to-one correspondence, the result of Step 3 is in $[s_{j,k-1},s_{j,k})$ if and only if the value on $R_S$ before Step 3 is in the image of $[s_{j,k-1},s_{j,k})$ under the map.

\subsubsection{implementation of respective gates}

We now consider how to implement respective gates in the PRN-on-a-register way to the level of four arithmetic operations.\\

\noindent (i) $V^{(j)}_k$

Note that most parts of $V^{(j)}_k$ consist of only arithmetic operations, addition, subtraction, multiplication, division and comparison, which mentioned in Sec \ref{sec:elemgate}.

\begin{figure*}[t]
	\begin{center}
		\includegraphics[width=1.2\textwidth,angle=270]{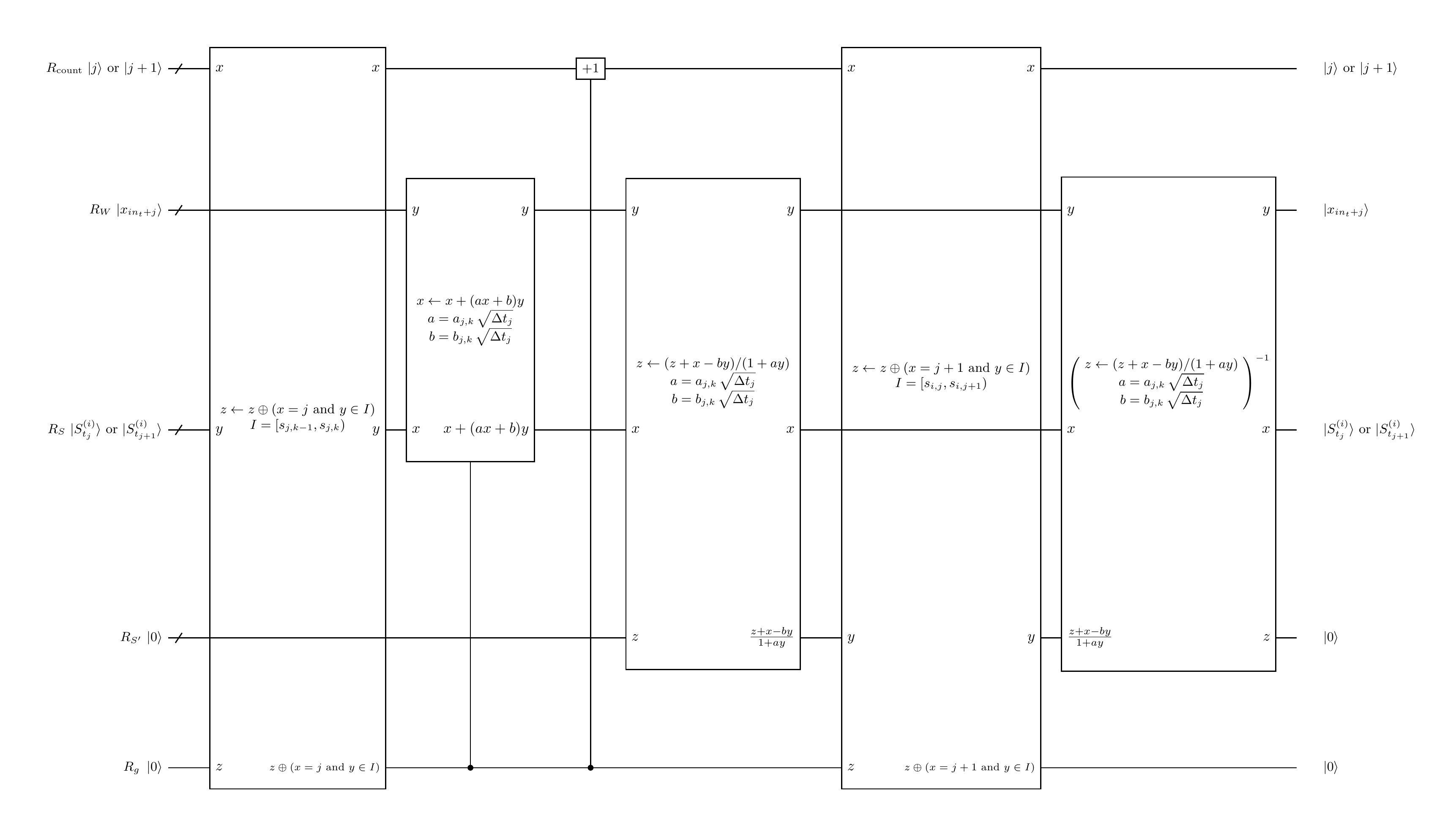}
	\end{center}
	\caption{$V^{(j)}_k$, which updates $R_S$ if the asset price is in the $k$-th grid of the LV function. Here and hereafter, the wire going over a gate means that the corresponding register is not used in the operation of the gate. A formula at the center of a gate represents the operation the gate performs and '-1' means the inverse operation. A formula beside a wire and in a gate means the input or the output of the gate.}
	\label{fig:Vjk}
\end{figure*}

\begin{figure*}[t]
	\begin{center}
		\includegraphics[width=1\textwidth]{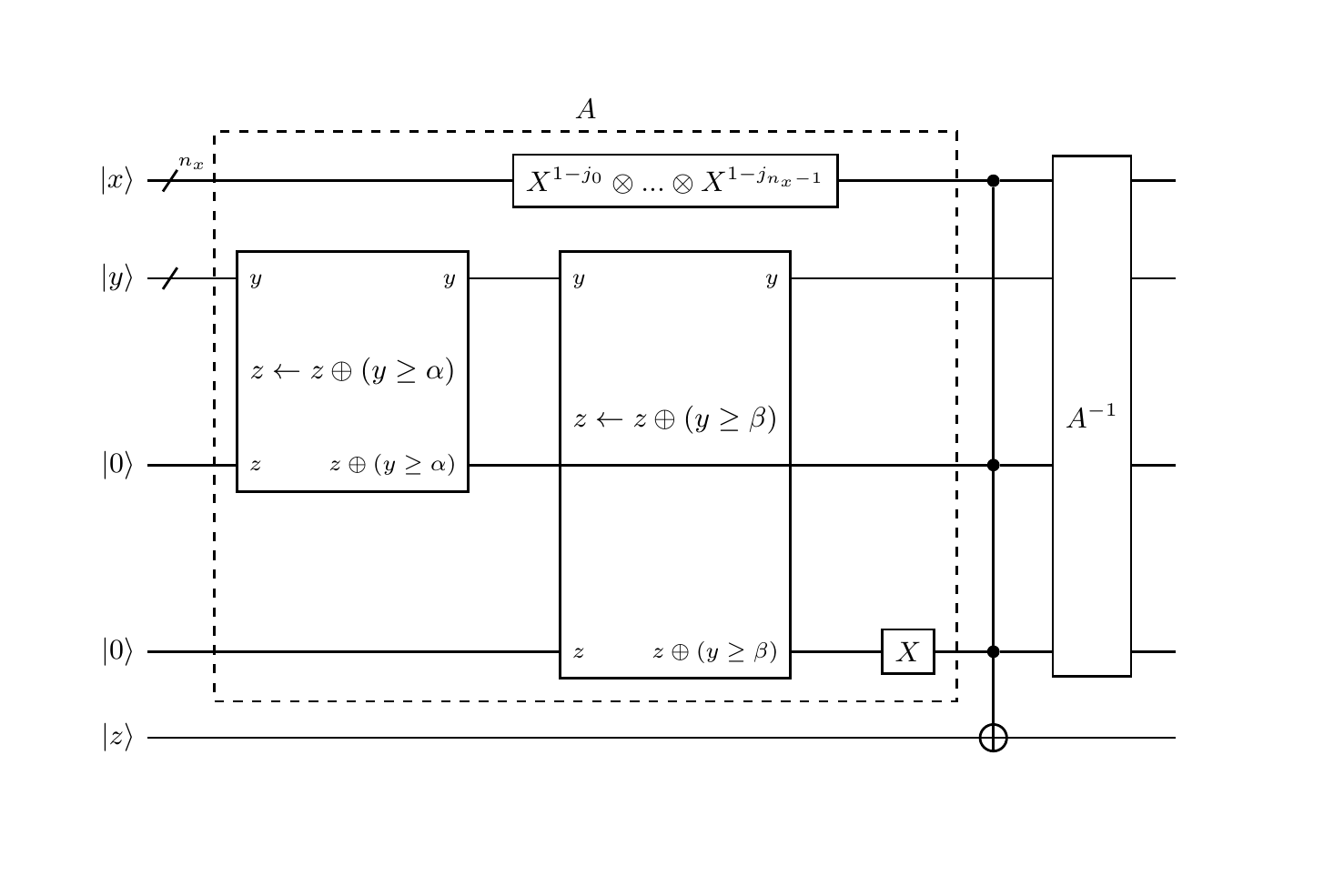}
	\end{center}
	\caption{The gate which outputs whether $x=j$ and $y\in [\alpha,\beta)$ or not. The control by a register means the multiple control by qubits therein.}
	\label{fig:comp}
\end{figure*}

For example, the gate $z\leftarrow z \oplus (x=j \ {\rm and} \ y\in I)$ can be divided to two parts.
The first part is checking that the value on $R_{\rm count}$ is equal to $j$ and this can be done by the multiple control Toffoli gate, which is studied in \cite{Selinger,Amy,Maslov}.
The second part is checking that the asset price is in a given interval, which can be constructed from two comparisons.
Combining these, the gate $z\leftarrow z \oplus (x=j \ {\rm and} \ y\in I)$ in Figure \ref{fig:Vjk} is constructed as shown in Figure \ref{fig:comp}.
Note that the bitwise flips $X^{1-j_0} \otimes ... \otimes X^{1-j_{n_x-1}}$ are operated before the multi control Toffoli.
Here, $j_a$ is the $a$-th digit of the binary representation of $j$, so the $a$-th qubit is flipped if and only if $j_a=0$.
This convert $\ket{x}$ to $\ket{1}...\ket{1}$ if and only if $x=j$.

The operation $x\leftarrow x+(ax+b)y$ in Figure \ref{fig:Vjk} can be realized as follows:
\begin{eqnarray}
\ket{x}\ket{y}\ket{0} & \rightarrow & \ket{x}\ket{y}\ket{1} \nonumber \\
 & \rightarrow & \ket{x}\ket{y}\ket{1+ay} \nonumber \\ 
& \rightarrow & \ket{(1+ay)x}\ket{y}\ket{1+ay} \nonumber \\
& \rightarrow & \ket{(1+ay)x+by}\ket{y}\ket{1+ay} \nonumber \\
& \rightarrow & \ket{(1+ay)x+by}\ket{y}\ket{0},
\label{eq:axby}
\end{eqnarray}
where the third ket corresponds to an ancillary register.
The first arrow is just setting a constant on a register.
The second arrow is the multiplication by a constant $a$.
The third arrow is the self-update multiplication, so it needs another ancillary register.
The fourth arrow is again multiplication by a constant $b$ and the final arrow is uncomputation of the first and second arrows.
Note that this is done under control by $R_g$.
In order for this to be controlled, it is sufficient to control only the second, fourth and final arrows, since the third arrow becomes a multiplication by 1 without the second.
Also note that multiplication by a $n$-bit constant $a$ or $b$ can be done by $n$ adders, that is, $n$ shift-and-add's: $ax=\sum_{i=0}^{n-1}{a_i2^ix}$, where $a_i$ is the $i$-th bit of $a$.
This saves qubits compared with the case where we use a multiplier, which needs holding $a$ on an ancillary register.

The operation $x\leftarrow (x-by)/(1+ay)$ in Figure \ref{fig:Vjk} is done as follows:
\begin{eqnarray}
\ket{x}\ket{y}\ket{0}\ket{0} & \rightarrow &  \ket{x}\ket{y}\ket{1}\ket{0} \nonumber \\ 
 & \rightarrow & \ket{x}\ket{y}\ket{1+ay}\ket{0} \nonumber \\ 
& \rightarrow & \ket{x-by}\ket{y}\ket{1+ay}\ket{0} \nonumber \\
& \rightarrow & \ket{x-by}\ket{y}\ket{1+ay}\ket{(x-by)/(1+ay)}, \label{eq:divInVjk}
\end{eqnarray}
where the first, second, third and fourth kets are $R_{S},R_{W}$, another ancillary register and $R_{S^\prime}$, respectively.
Here, the first and second arrows are same as (\ref{eq:axby}), the third is the multiplication by a constant $-b$ and the final one is division.
Here, we do not have to uncompute $R_{S}$ and the ancillary register, since the whole of this operation is uncomputed soon later in $V_{j,k}$.
\\

\noindent (ii) $J_W, P_W$

\begin{figure*}[t]
	\begin{minipage}{1\hsize}
		\begin{center}
			\includegraphics[width=0.4\textwidth]{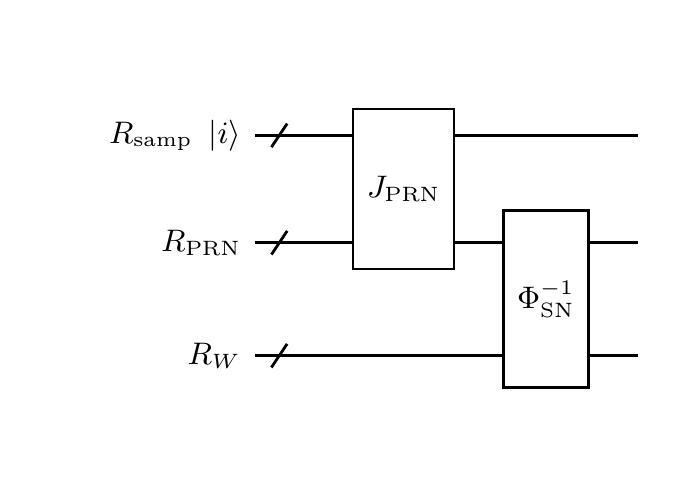}
		\end{center}
		\subcaption{$J_W$, the gate to let the PSNRN sequence jump to the $in_t+1$ element.}
	\end{minipage}	
	
	\begin{minipage}{1\hsize}
		\begin{center}
			\includegraphics[width=0.5\textwidth]{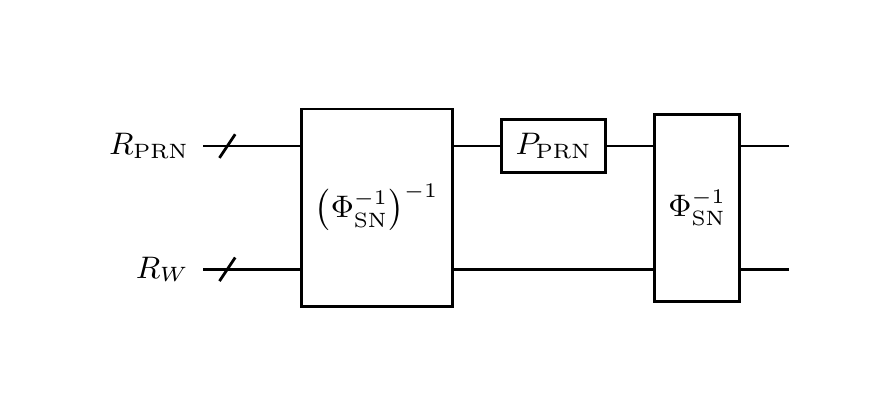}
		\end{center}
		\subcaption{$P_W$, the gate to progress the PSNRN sequence by a step.}
	\end{minipage}	
	\caption{The circuits to generate PSNRN sequences.}
	\label{fig:JWPW}
\end{figure*}

In \cite{Miyamoto}, implementation of PRN on quantum circuits is presented, using the PRN generator called PCG\cite{PCG}.
Note that this PRN generator generates {\it uniform} RNs.
We now need PSNRNs, so we must transform uniform distribution to standard normal distribution.
There are some method and we adopt the inverse transform sampling.
Schematically, $J_W$ and $P_W$ are implemented as shown in Figure \ref{fig:JWPW}.
Here, $J_{PRN}$ is the gate to let the PRN sequence jump to the $in_t+1$ and $P_{PRN}$ is the gate to progress the PRN sequence by a step.
They sequentially generate uniform RNs on the ancillary register $R_{\rm PRN}$ and they are transformed to PSNRN on $R_W$.

Although we refer to \cite{Miyamoto} for the detail of implementation of the PRN generator, we here briefly explain.
This generator is combination of linear congruential generator (LCG) and permutation of bit string.
For LCG, update of the PRN sequence is done by
\begin{equation}
x_{n+1} = ax_n + c \ {\rm mod} \ N, \label{eq:LCGProg}
\end{equation}
where $a,N$ are positive integers and $c$ is a nonnegative integer, and the $n$-th element of the sequence is computed from $n$ and the initial value $x_0$ by
\begin{equation}
x_n = a^nx_0 + \frac{c(a^n-1)}{a-1} \ {\rm mod} \ N, \label{eq:LCGJump}.
\end{equation}
According to \cite{Vedral}, we can construct the modular adder using 5 plain adders.
Modular multiplication by a $n$-bit constant can be done as $n$ modular shift-and-add's.
Modular division by a constant $a-1$ can be done as modular multiplication by a constant integer $\beta$ such that $\beta(a-1)=1 \ {\rm mod} \ N$, if exists.
Modular exponentiation $a^x \ {\rm mod} \ N$ is computed as a sequence controlled modular multiplication\cite{Vedral}.
So, to summarize, we can perform (\ref{eq:LCGProg}) and (\ref{eq:LCGJump}) using only controlled adders.
In (\ref{eq:LCGProg}), the state is transformed as
\begin{eqnarray}
\ket{x}\ket{0}\ket{0} & \rightarrow & \ket{x}\ket{ax \ {\rm mod} \ N}\ket{0} \nonumber \\
& \rightarrow & \ket{0}\ket{ax \ {\rm mod} \ N}\ket{0} \nonumber \\
& \rightarrow & \ket{0}\ket{ax \ {\rm mod} \ N}\ket{c} \nonumber \\
& \rightarrow & \ket{0}\ket{ax+c \ {\rm mod} \ N}\ket{c} \nonumber \\
& \rightarrow & \ket{0}\ket{ax+c \ {\rm mod} \ N}\ket{0}. \label{eq:ProgTransf}
\end{eqnarray}
In the circuit we are considering, the first and second registers are $R_{\rm PRN}$ and an ancillary register, which interchange their role at every step, and the third is another ancillary register.
Each step corresponds to an elementary operation as follows.
The first arrow is modular multiplication.
The second arrow is the inverse modular multiplication by a integer $\alpha$ such that $a\alpha = 1 \ {\rm mod} \ N$ and this clearing step is necessary to avoid increase of ancillas.
The third is just loading $c$ on an ancillary register, the fourth is modular addition and the last is unloading.
(\ref{eq:LCGJump}) progresses as follows:
\begin{eqnarray}
& & \ket{n}\ket{0}\ket{0}\ket{0} \nonumber \\
& \rightarrow & \ket{n}\ket{a^n \ {\rm mod} \ N}\ket{0}\ket{0} \nonumber \\
& \rightarrow & \ket{n}\ket{a^n \ {\rm mod} \ N}\ket{\left(x_0+\frac{c}{a-1}\right)a^n \ {\rm mod} \ N}\ket{0} \nonumber \\
& \rightarrow & \ket{n}\ket{a^n \ {\rm mod} \ N}\ket{\left(x_0+\frac{c}{a-1}\right)a^n \ {\rm mod} \ N}\ket{\frac{c}{a-1}} \nonumber \\
& \rightarrow & \ket{n}\ket{a^n \ {\rm mod} \ N}\ket{\left(x_0+\frac{c}{a-1}\right)a^n - \frac{c}{a-1} \ {\rm mod} \ N}\ket{\frac{c}{a-1}} \nonumber \\
& \rightarrow &  \ket{n}\ket{0}\ket{\left(x_0+\frac{c}{a-1}\right)a^n - \frac{c}{a-1} \ {\rm mod} \ N}\ket{0}, \label{eq:JumpTransf}
\end{eqnarray}
where the first, third, second and fourth registers are respectively $R_{\rm samp}, R_{\rm PRN}$ and two ancillary registers.
The first arrow is modular exponentiation, the second is modular multiplication, the third is loading, the fourth is modular addition and the last is uncomputation of the first and third.

We do not explain permutation: see \cite{Miyamoto} for the detail.
We just make a comment that it is implemented by a simple circuit, for example, Xorshift is implemented as a sequence of CNOT.\\

We also need the gate to calculate $\Phi_{\rm SN}^{-1}$, the inverse function of CDF of standard normal distribution.
There are some ways to calculate this efficiently and we here adopt the method in \cite{Hormann}.
In the method, $\mathbb{R}$ is divided into some intervals and $\Phi_{\rm SN}^{-1}$ is approximated by a polynomial in each interval.
We adopt the setting where the number of intervals is 109\footnote{if we include $(-\infty,x^{\rm ICDF}_0)$ and $[x^{\rm ICDF}_{n^{\rm ICDF}},\infty,)$, it is 111} and polynomials are cubic, which realizes the error smaller than $10^{-6}$.
Here, we write the approximated inverse CDF as
\begin{equation}
\Phi_{\rm SN}^{-1}(x) \approx c_{m,3}x^3+c_{m,2}x^2+c_{m,1}x+c_{m,0}
\end{equation}
for $x^{\rm ICDF}_{m-1}\le x< x^{\rm ICDF}_{m},m=0,1,...,n_{\rm ICDF}+1$, where $x^{\rm ICDF}_0<x^{\rm ICDF}_1<...<x^{\rm ICDF}_{n_{\rm ICDF}}$ are the end points of the intervals and $n_{\rm ICDF}$ is the number of the interval.
Consider that $x^{\rm ICDF}_{-1}=-\infty,x^{\rm ICDF}_{n_{\rm ICDF}+1}=+\infty$.

\begin{figure*}[t]
	\begin{center}
		\includegraphics[width=1.3\textwidth,angle=270]{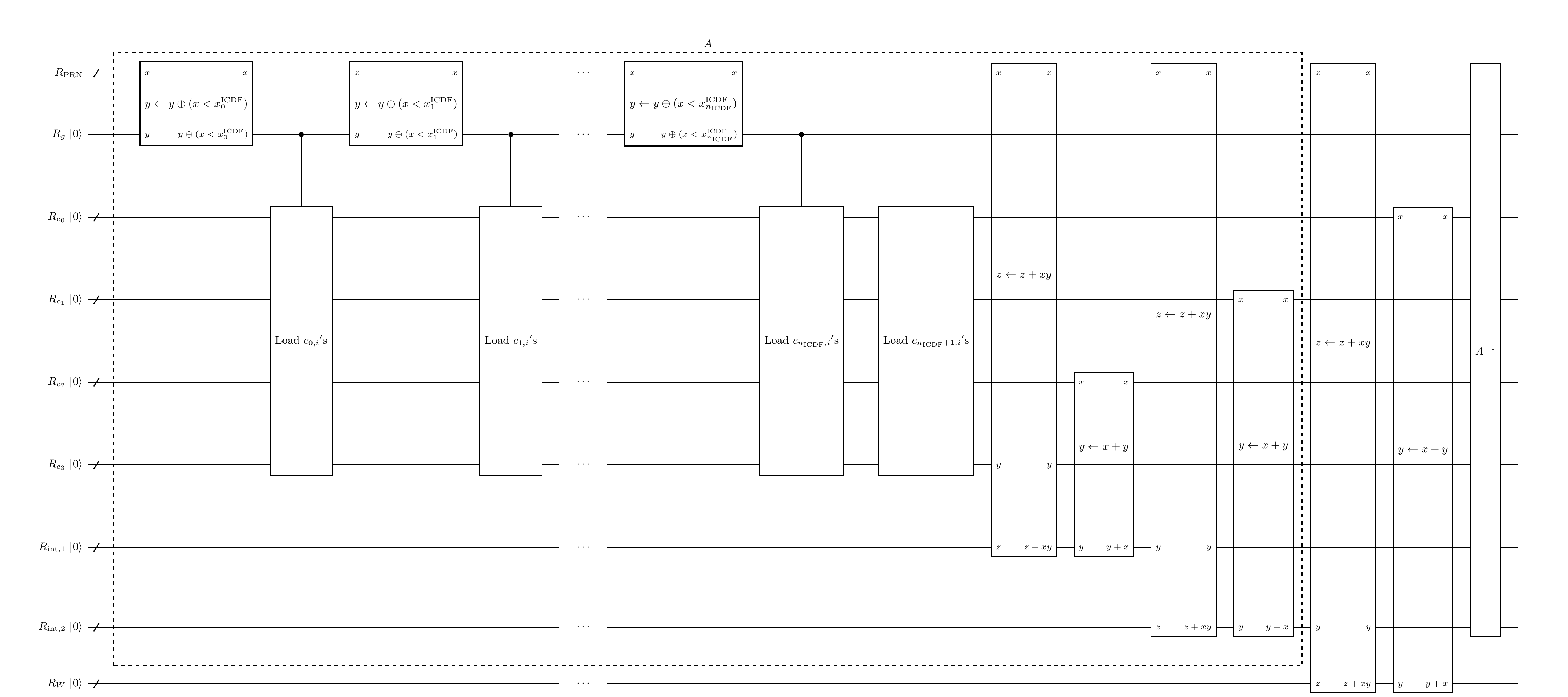}
	\end{center}
	\caption{The gate to calculate the inverse CDF of standard normal distribution by piecewise polynomial approximation.}
	\label{fig:InvCDF}
\end{figure*}

\begin{figure*}[t]	
	\begin{minipage}{1\hsize}
		\begin{center}
			\includegraphics[width=0.5\textwidth]{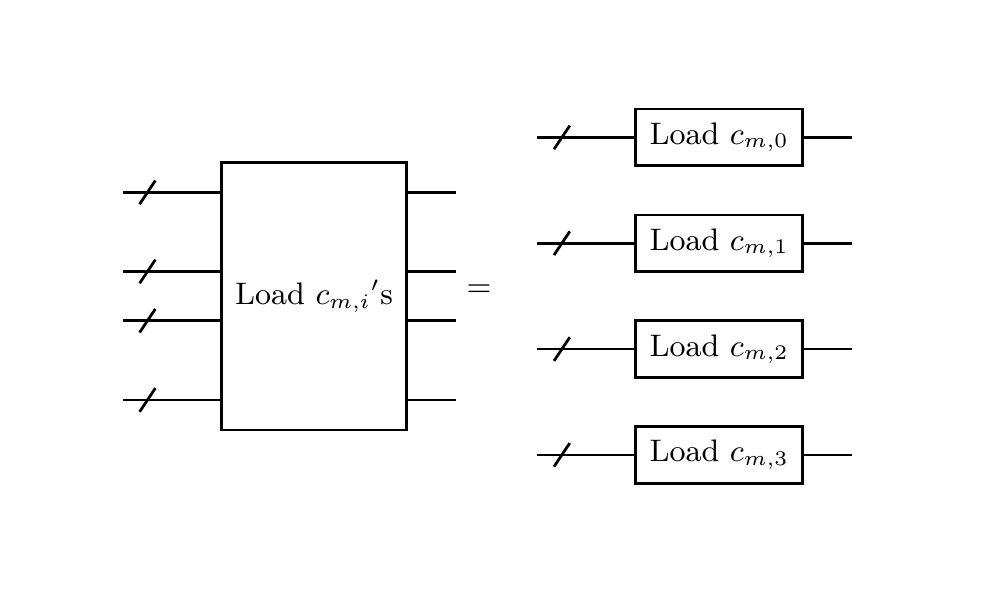}
		\end{center}
		\subcaption{The detail of the ``Load $c_{m,i}$'s" gate.}
	\end{minipage}
	
	\begin{minipage}{1\hsize}
		\begin{center}
			\includegraphics[width=0.5\textwidth]{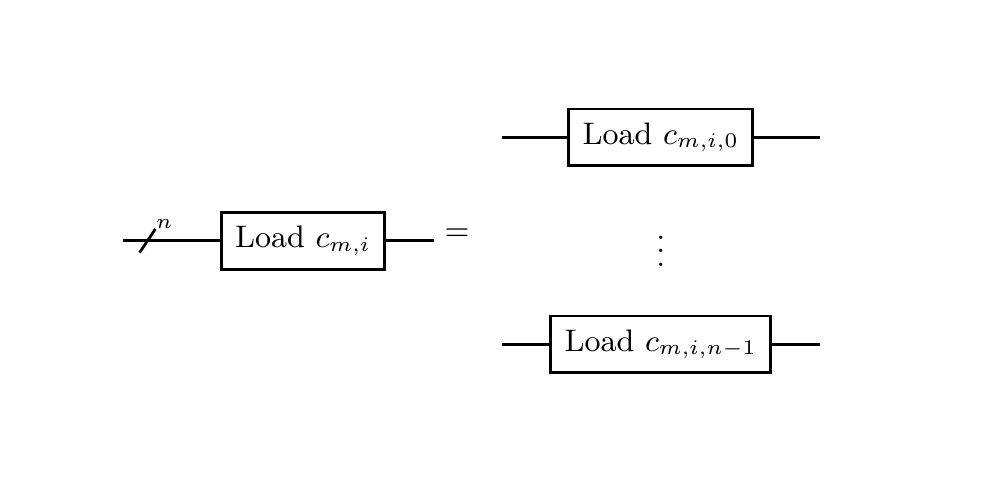}
		\end{center}
		\subcaption{The detail of the ``Load $c_{m,i}$" gate.}
	\end{minipage}
	
	\begin{minipage}{1\hsize}
		\begin{center}
			\includegraphics[width=1\textwidth]{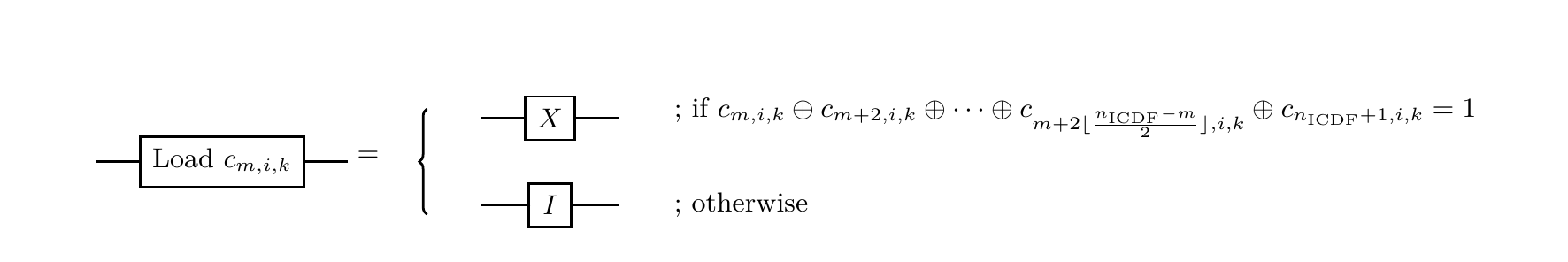}
		\end{center}
		\subcaption{The detail of the ``Load $c_{m,i,k}$" gate.}
	\end{minipage}
	
	\caption{Parts of the circuit in Figure \ref{fig:InvCDF}. Here, $c_{m,i,k}$ means the $k$-th digit of $c_{m,i}$.}
	\label{fig:PartsInvCDF}
\end{figure*}

Such a piecewise cubic function can be implemented as Figure \ref{fig:InvCDF}.
We here explain how it works.
First, the sequence of comparators and ``Load $c_{m,i}'s$" gates load $c_{m,0},...,c_{m,3}$ into the register $R_{c_0},...,R_{c,3}$ respectively as follows.
The comparators compare the value $x$ of $R_{\rm PRN}$ and the grid points $x^{\rm ICDF}_m$ and flip $R_g$ if $x<x^{\rm ICDF}_m$.
If $R_g$ is 1, the ``Load $c_{m,i}'s$" gates are activated.
They are actually collections of bitwise flips, that is, $X$ gates.
If $x \ge x^{\rm ICDF}_{n_{\rm ICDF}}$, only ``Load $c_{n_{\rm ICDF}+1,i}'s$" gate is performed and it loads $c_{n_{\rm ICDF}+1,0},...,c_{n_{\rm ICDF}+1,3}$.
If $x^{\rm ICDF}_{n_{\rm ICDF}-1} \le x < x^{\rm ICDF}_{n_{\rm ICDF}}$, ``Load $c_{n_{\rm ICDF},i}'s$" and ``Load $c_{n_{\rm ICDF}+1,i}'s$" are performed.
So, we set ``Load $c_{n_{\rm ICDF},i}'s$" so that it compensates flips done by ``Load $c_{n_{\rm ICDF}+1,i}'s$" and $c_{n_{\rm ICDF},0},...,c_{n_{\rm ICDF},3}$ are successfully loaded.
The case where $x$ is in another interval is similar.
The point is that if $x^{\rm ICDF}_{M-1}\le x< x^{\rm ICDF}_{M}$, every other gates are activated.
That is, the activated gates are ``Load $c_{m,i}'s$" of $m=M,M+2,...,n_{\rm ICDF},n_{\rm ICDF}+1$ if $n_{\rm ICDF}-M$ is even and $m=M,M+2,...,n_{\rm ICDF}-1,n_{\rm ICDF}+1$ if $n_{\rm ICDF}-M$ is odd.
This is because every comparator after the $M$-th one flips $R_g$ and $R_g$ takes 0 and 1 alternatingly. 
Considering this, the $X$ gates in ``Load $c_{m,i}'s$" are set as Figure \ref{fig:PartsInvCDF}, so that $c_{m,0},...,c_{m,3}$ for appropriate $m$ are loaded after the sequence of all activated gates.
After load of coefficients, the cubic function is calculated in the Horner's method
\begin{equation}
((c_{m,3}x + c_{m,2})x + c_{m,1})x + c_{m,0}.
\end{equation}
This is done by the sequence of adders and multipliers in the latter half of the circuit in Figure \ref{fig:InvCDF}.\\

\noindent (iii) Payoff

In this paper, we do not consider gates to calculate payoffs in detail, since the resource the gates require is same in both the PRN-on-a-register way and the register-per-RN way.
We here make just a short comment.
In many cases, a payoff can be expressed in the following form:
\begin{equation}
f^{\rm pay}_{i}=\min \{\max \{a_iS_{t_i}+b_i,f_i\},c_i\}, \label{eq:payCapFloor}
\end{equation}
where $a_i,b_i,c_i,f_i$ are real constants, that is, a linear function of the asset price with the upper bound ({\it cap}) $c_i$ and the lower bound ({\it floor}) $f_i$.
For example, a payoff in an European call option (\ref{eq:PayECall}) corresponds to $a_i=1,b_i=-K,c_i=+\infty,f_i=0$.
Payoffs expressed as (\ref{eq:payCapFloor}) can be calculated easily by combination of comparators, adders and multipliers.

\subsection{the register-per-RN way}

\subsubsection{calculation flow}

Also for the register-per-RN way, we start from presenting the calculation flow, which is somewhat simpler than the PRN-on-a-register way.
Again, we consider the flow until calculation of the payoff sum.

Before we present the calculation flow, we explain the required registers.
\begin{itemize}
	
	\item $R_{W_i},i=1,...,n_t$
	
	This is a register for the $i$-th SNRN.
	We need such a register per random number, so the total number is $n_t$.
	
	\item $R_{S_i},i=0,1,...,n_t$
	
	This is a register where the value of the asset price at time $t_i$ is held.
	
	\item $R_{{\rm payoff},i},i=1,...,n_t$
	
	This is a register where the value of the sum of payoffs by $t_i$ is held.
	
\end{itemize}
We again omit ancillary registers here and explain them later.
Besides, we again assume that these and ancillary registers necessary to update them have $n_{\rm dig}$ qubits.

We here concretely define a superposition of SNRN values as the following state.
In advance, we set the equally spaced $N_{\rm SN}+1$ points for discretization of the distribution $x_{{\rm SN},0}<x_{{\rm SN},1}<...<x_{{\rm SN},N_{\rm SN}}$, where $x_{{\rm SN},0}$ and $x_{{\rm SN},N_{\rm SN}}$ are the upper and lower bounds of the distribution and set to, say, -4 and +4, respectively.
We here assume $N_{\rm SN}=2^{n_{\rm dig}}$ for simplicity.
Then, we define $\ket{\rm SN}$ as 
\begin{equation}
\ket{\rm SN} = \sum_{i=0}^{N_{\rm SN}-1}{\sqrt{p_{{\rm SN},i}}\ket{i}},
\end{equation}
where $p_{{\rm SN},i}=\int_{x_{{\rm SN},i}}^{x_{{\rm SN},i+1}}\phi_{\rm SN}(x)dx$ and $\phi_{\rm SN}(x)$ is the probability density function of the standard normal distribution.
We consider how to create such a state later.
Since $x_{{\rm SN},i}$ can be easily calculated from the index $i$ by a linear function, we identify $i$ as $x_{{\rm SN},i}$.

Then, the calculation flow is as follows:
\begin{enumerate}
	\item Initialize all registers to $\ket{0}$ except $R_{S_0}$, which is initialized to $\ket{S_{t_0}}$.
	\item Generate superpositions of SNRNs on $R_{W_1},...,R_{W_{n_t}}$. That is, set each of them to $\ket{\rm SN}$.
	\item Perform the time evolution (\ref{eq:EM}) using the value on $R_{W_1}$ as $w_1$. The result is output to $R_{S_1}$ as $\ket{S_{t_1}}$.
	\item Calculate the payoff at time $t_1$ using $R_{S_1}$ and output the sum of it and the previous payoffs to $R_{{\rm payoff},i}$.
	\item Iterate operations similar to 3-4 for each time step until the time reaches $t_{n_t}$.
	\item Finally we obtain a superposition of states in which the value on $R_{{\rm payoff},n_t}$ is the sum of payoffs for each pattern of values of SNRNs. Estimate the expectation value of $R_{{\rm payoff},n_t}$ to get (\ref{eq:priceMCRN}).
\end{enumerate}

The flow of state transformation is as follows.
Writing only $R_{W_1},...,R_{W_{n_t}}$, $R_{S_0},R_{S_1},...,R_{S_{n_t}}$ and $R_{{\rm payoff},1},...,R_{{\rm payoff},n_t}$,
\begin{widetext}
\begin{eqnarray}
& &   \ket{0}^{\otimes n_t}\ket{S_{t_0}}\ket{0}^{\otimes n_t}\ket{0}^{\otimes n_t} \nonumber \\ 
& \xrightarrow{2} &\ket{\rm SN}^{\otimes n_t}\ket{S_{t_0}}\ket{0}^{\otimes n_t}\ket{0}^{\otimes n_t} \nonumber \\
& \xrightarrow{3} &\sum_{i_1=0}^{N_{\rm SN}-1}\sqrt{p_{{\rm SN},i_1}}\ket{i_1}\ket{\rm SN}^{\otimes n_t-1}\ket{S_{t_0}}\ket{S^{(i_1)}_{t_1}}\ket{0}^{\otimes n_t-1}\ket{0}^{\otimes n_t} \nonumber \\
& \xrightarrow{4} & \sum_{i_1=0}^{N_{\rm SN}-1}\sqrt{p_{{\rm SN},i_1}}\ket{i_1}\ket{\rm SN}^{\otimes n_t-1}\ket{S_{t_0}}\ket{S^{(i_1)}_{t_1}}\ket{0}^{\otimes n_t-1}\ket{f^{\rm pay}_{i_1}(S^{(i_1)}_{t_1})}\ket{0}^{\otimes n_t-1} \nonumber \\
& \xrightarrow{5} & ...\nonumber \\
& \xrightarrow{5} &\sum_{i_1,\cdots,i_{n_t}=0}^{N_{\rm SN}-1}\sqrt{p_{{\rm SN},i_1}...p_{{\rm SN},i_{n_t}}}\ket{i_1}...\ket{i_{n_t}}\ket{S_{t_0}}\ket{S^{(i_1)}_{t_1}}...\ket{S^{(i_1\cdots i_{n_t})}_{t_{n_t}}}\ket{f^{\rm pay}_{i_1}(S^{(i_1)}_{t_1})}...\ket{\sum_{j=1}^{n_t}{f^{\rm pay}_{j}(S^{(i_1...i_j)}_{t_j})}},
    \label{eq:calcflowRN}
\end{eqnarray}
\end{widetext}
where $S^{(i_1...i_j)}_{t_j}$ is the value of the asset price at time $t_j$ evolved by $w_1=x_{{\rm SN},i_1},...,w_j=x_{{\rm SN},i_j}$.

\subsubsection{overview of the circuit}
\begin{figure*}[t]
	\begin{center}
		\includegraphics[width=0.7\textwidth]{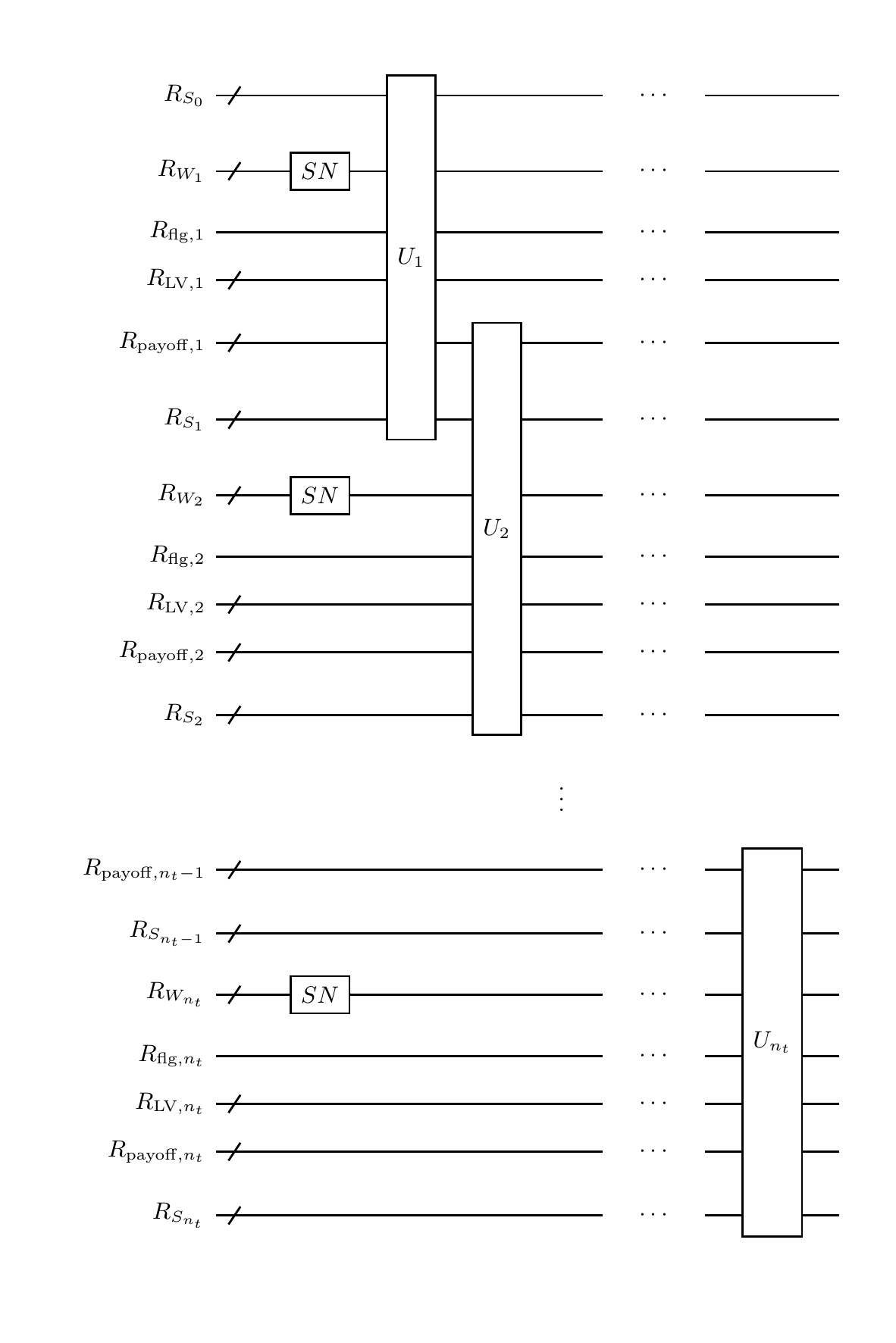}
	\end{center}
	\caption{The overview of the circuit for asset price evolution in the LV model in the register-per-RN way.}
	\label{fig:overviewRN}
\end{figure*}

\begin{figure*}[t]
	\begin{center}
		\includegraphics[width=1.2\textwidth,angle=270]{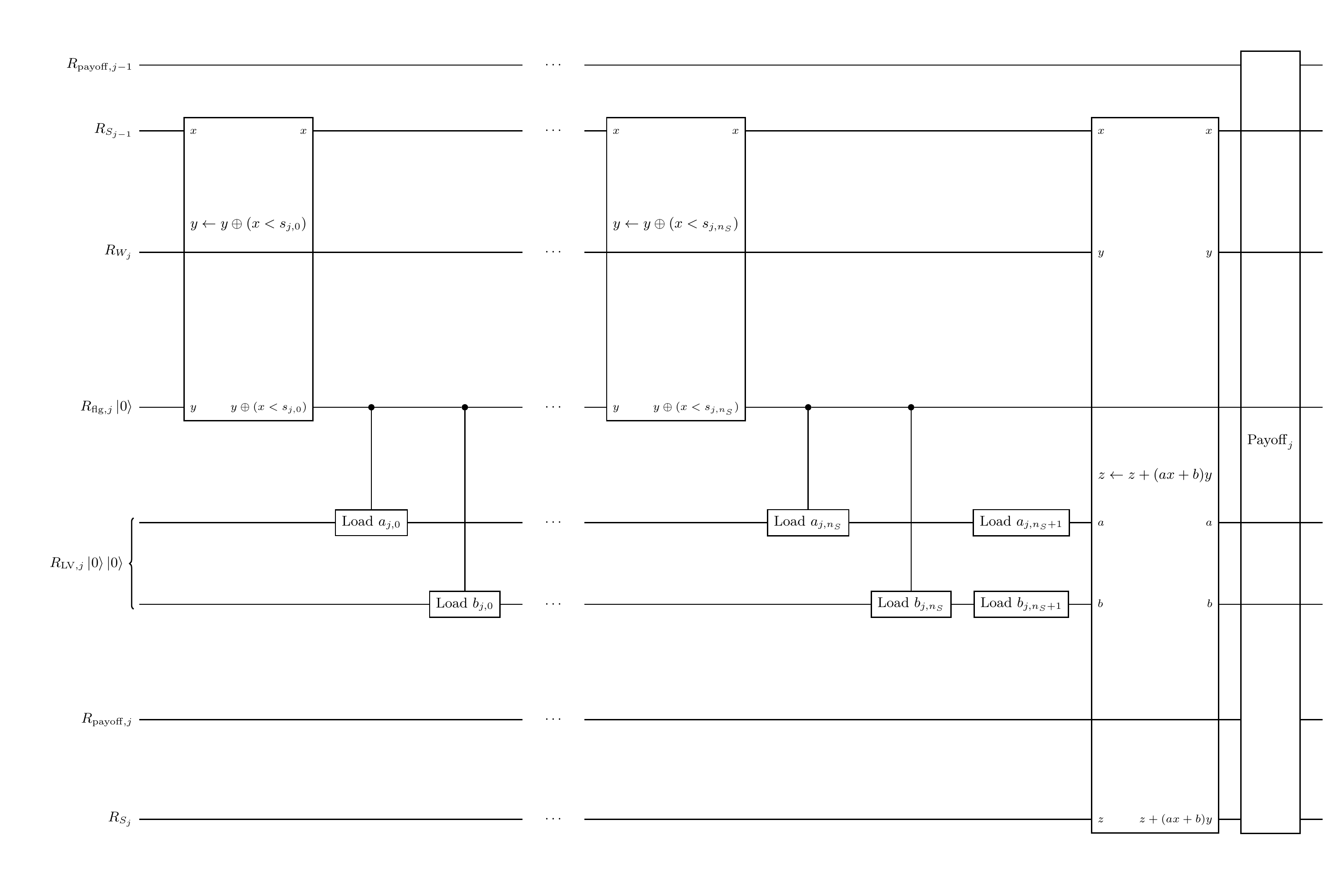}
	\end{center}
	\caption{$U_j$, which performs the $j$-th step of asset price evolution, in the register-per-RN way.}
	\label{fig:UtRN}
\end{figure*}

The outline of the circuit in the register-per-RN is as shown in Figure \ref{fig:overviewRN}.
First, $\ket{\rm SN}$ is created on each $R_{W_j}$ by the gate SN, which is considered in detail later.
After that, the gate $U_j$ performs $j$-th step of asset price evolution and payoff calculation.
For each evolution step, ancillary registers $R_{{\rm flg},j}$ and $R_{{\rm LV},j}$, which have 1 and $2n_{\rm dig}$ qubits respectively, are necessary.
$U_j$ is then implemented as Figure \ref{fig:UtRN}.
In this gate, the sequence of comparators and ``Load" gates set $a_{j,k},b_{j,k}$ in (\ref{eq:LV}) into $R_{{\rm LV},j}$ by the trick similar to that in the circuit in Figure \ref{fig:InvCDF}.
Then, the operation $x\leftarrow x+(ax+b)y$ updates the asset price according to (\ref{eq:EM}).
Since the register-per-RN way does not aim to uncompute ancillary qubits in order to reduce them, $x\leftarrow x+(ax+b)y$ can be done as follows:
\begin{eqnarray}
\ket{x}\ket{y}\ket{a}\ket{b}\ket{0}\ket{0} & \rightarrow & \ket{x}\ket{y}\ket{a}\ket{b}\ket{0}\ket{x} \nonumber \\ 
& \rightarrow &\ket{x}\ket{y}\ket{a}\ket{b}\ket{xy}\ket{x} \nonumber \\
& \rightarrow & \ket{x}\ket{y}\ket{a}\ket{b}\ket{xy}\ket{x+axy} \nonumber \\
& \rightarrow & \ket{x}\ket{y}\ket{a}\ket{b}\ket{xy}\ket{x+axy+by}, \label{eq:axbyRN}
\end{eqnarray}
where the first ket is $R_{S_{j-1}}$, the second is $R_{W_j}$, the third and fourth are $R_{{\rm LV},j}$, the fifth is an ancillary register and the last is $R_{S_{j}}$.
So, this operation consists of copying a state and three multiplications.
At the end of $U_j$, the payoff is calculated using $R_{S_{j}}$.
Here, the "${\rm Payoff}_j$" gate performs the following operation
\begin{eqnarray}
& & \ket{S^{(i_1\cdots i_j)}_{t_j}}\ket{\sum_{k=1}^{j-1}{f^{\rm pay}_{k}(S^{(i_1...i_k)}_{t_k})}}\ket{0} \nonumber \\
&\rightarrow& \ket{S^{(i_1\cdots i_j)}_{t_j}}\ket{\sum_{k=1}^{j-1}{f^{\rm pay}_{k}(S^{(i_1...i_k)}_{t_k})}}\ket{\sum_{k=1}^{j}{f^{\rm pay}_{k}(S^{(i_1...i_k)}_{t_k})}},
\end{eqnarray}
where the first, second and third ket are $R_{S_{j}}$, $R_{{\rm payoff},j-1}$ and $R_{{\rm payoff},j}$.
This is done as copying $R_{{\rm payoff},j-1}$ to $R_{{\rm payoff},j}$ followed by calculation and addition of $f^{\rm pay}_{j}(S^{(i_1...i_j)}_{t_j})$ to $R_{{\rm payoff},j}$.

\subsubsection{implementation of the SN gate}

Let us now consider implementation of the SN gate, which creates a superposition of SNRN values.
The outline was presented in \cite{Grover}.
In addition to this, we here explain some details, which were not explicitely explained in \cite{Grover}.

We construct the state in the recursive way.
We assume that we have already divided $[x_{{\rm SN},0},x_{{\rm SN},N_{\rm SN}}]$ by equally-spaced $2^m+1$ points $x_{{\rm SN},0}^{(m)}=x_{{\rm SN},0}<x^{(m)}_{{\rm SN},1}<...<x^{(m)}_{{\rm SN},2^m}=x_{{\rm SN},N_{\rm SN}}$ and created
\begin{equation}
\ket{{\rm SN}_m} = \sum_{i=0}^{2^m-1}{\sqrt{p^{(m)}_{{\rm SN},i}}\ket{i}},
\end{equation}
where $p^{(m)}_{{\rm SN},i}=\int_{x^{(m)}_{{\rm SN},i}}^{x^{(m)}_{{\rm SN},i+1}}\phi_{\rm SN}(x)dx$.
We also assume that we have a gate to efficiently compute $\theta^{(m)}_i=\arccos\sqrt{f^{(m)}_i}$ with the input $i$, where $f^{(m)}_i$ is 
\begin{equation}
f^{(m)}_i=\frac{\int_{x^{(m)}_{{\rm SN},i}}^{\left(x^{(m)}_{{\rm SN},i}+x^{(m)}_{{\rm SN},i+1}\right)/2}\phi_{\rm SN}(x)dx}{\int_{x^{(m)}_{{\rm SN},i}}^{x^{(m)}_{{\rm SN},i+1}}\phi_{\rm SN}(x)dx}.
\end{equation}
Then, the following state transformation is possible:
\begin{eqnarray}
\ket{{\rm SN}_m}\ket{0}\ket{0} & = & \sum_{i=0}^{2^m-1}{\sqrt{p^{(m)}_{{\rm SN},i}}\ket{i}\ket{0}\ket{0}} \nonumber \\
& \rightarrow & \sum_{i=0}^{2^m-1}{\sqrt{p^{(m)}_{{\rm SN},i}}\ket{i}\ket{0}\ket{\theta^{(m)}_i}} \nonumber \\
& \rightarrow & \sum_{i=0}^{2^m-1}{\sqrt{p^{(m)}_{{\rm SN},i}}\ket{i}\left(\cos\theta^{(m)}_i\ket{0}+\sin\theta^{(m)}_i\ket{1}\right)\ket{\theta^{(m)}_i}} \nonumber \\
& = & \sum_{i=0}^{2^{m+1}-1}{\sqrt{p^{(m+1)}_{{\rm SN},i}}\ket{i}\ket{\theta^{(m)}_i}} \nonumber \\
& = & \ket{{\rm SN}_{m+1}}\ket{0},
\end{eqnarray}
where we use the gate to compute $\theta^{(m)}_i$ at the first arrow and perform the controlled rotation at the second arrow. 
Repeating this until $m=n_{\rm dig}-1$, we get $\ket{\rm SN}$.

However, as far as the authors know, neither \cite{Grover} nor other papers present the gate to compute $\theta^{(m)}_i$.
Here, we propose a way to do this on the basis of a simple Taylor expansion.
Consider
\begin{equation}
g(x,\delta)=\frac{\int_x^{x+\delta/2}\phi_{\rm SN}(x)dx}{\int_x^{x+\delta}\phi_{\rm SN}(x)dx}.
\end{equation}
We can show that
\begin{equation}
g(x,\delta)\approx \frac{1}{2} + \frac{1}{8}\delta x + \frac{1}{16}\delta^2 + \mathcal{O}(\delta^3)
\end{equation}
by simple calculation.
So, $g(x,\delta)$ is well-approximated by a linear function of $x$ for small $\delta$.
We can use this to compute $f^{(m)}_i$, since
\begin{equation}
f^{(m)}_i=g\left(x^{(m)}_{{\rm SN},i},\frac{\Delta}{2^m}\right), \Delta=x_{{\rm SN},N_{\rm SN}}-x_{{\rm SN},0}.
\end{equation}
Given $x_{{\rm SN},N_{\rm SN}},x_{{\rm SN},0}$ and $m$ large enough that $\Delta/2^m$ is sufficiently small, this can be approximately seen as a linear function of $i$, since $x^{(m)}_{{\rm SN},i}=x_{{\rm SN},0}+\frac{\Delta}{2^m}i$.
Actually, we numerically confirmed that for $m\ge 7$ the above approximation gives error smaller than $10^{-5}$.

\begin{figure*}[t]

	\begin{minipage}{1\hsize}
		\begin{center}
			\includegraphics{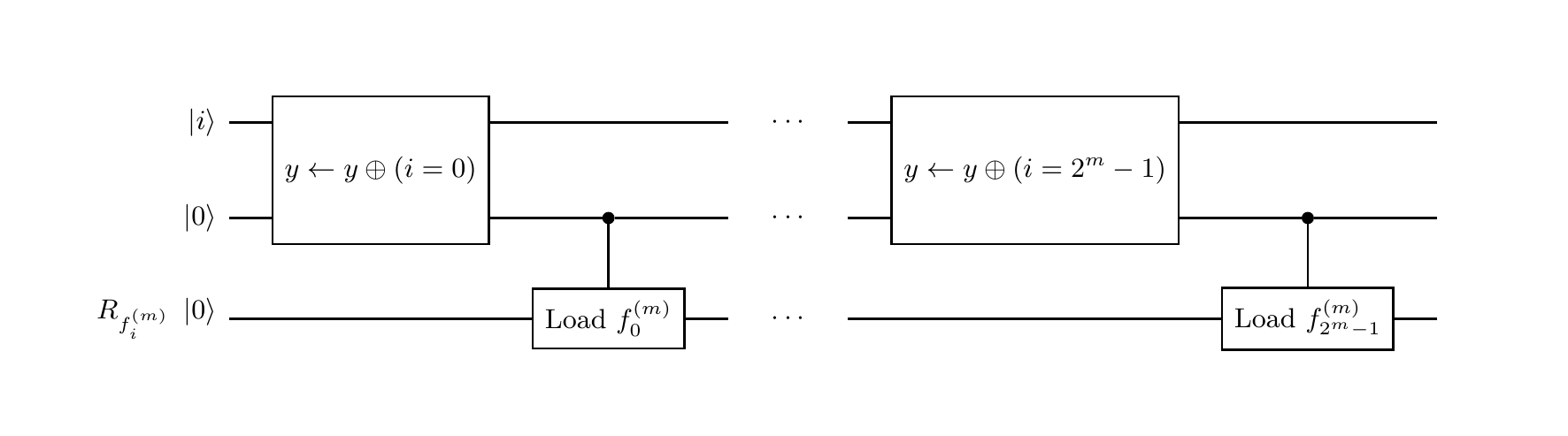}
		\end{center}
		\subcaption{For $m\le 6$.}
		\label{fig:fmismallm}
	\end{minipage}
	
	\begin{minipage}{1\hsize}
		\begin{center}
			\includegraphics{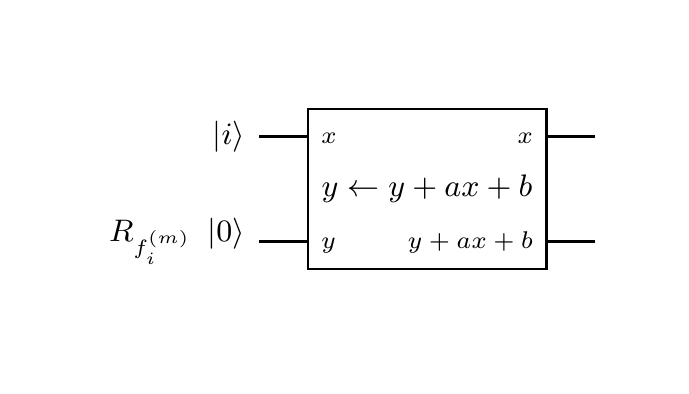}
		\end{center}
		\subcaption{For $m\ge 7$.}
		\label{fig:fmilargem}
	\end{minipage}
	
	\caption{Circuit to compute $f^{(m)}_i$.}
	\label{fig:fmi}
\end{figure*}

We then reach the circuit in Figure \ref{fig:fmi} for calculation of $f^{(m)}_i$.
For $m\le 6$, as shown in Figure \ref{fig:fmismallm}, the sequence of comparators and ``Load" gates set $f^{(m)}_i$ to the output register $R_{f^{(m)}_i}$, using the trick similar to the circuit in Figure \ref{fig:InvCDF} again.
Note that comparators now check whether the input value $i$ is equal to a given integer constant or not, so each of them is actually a combination of bitwise flips and a multi-controlled Toffoli gate, similar to that appeared in Figure \ref{fig:comp}.
Also note that, if the input $i$ is $I$, ``Load $f^{(m)}_i$" gates are activated for $i\ge I$.
Considering this point, each ``Load" gate is set so that operations of it and the following gates are successfully compensated.
For $m\ge 7$, the $f^{(m)}_i$ is just a linear transformation, which is implemented as bitwise flips followed by a constant multiplier.
Of course, according to required accuracy, the value of $m$ where the two types of $f^{(m)}_i$ are switched should be adjusted and the degree of the Taylor approximation for $g(x,\delta)$ should be increased.

\begin{figure*}[t]	
	
	\begin{minipage}{1\hsize}
		\begin{center}
			\includegraphics{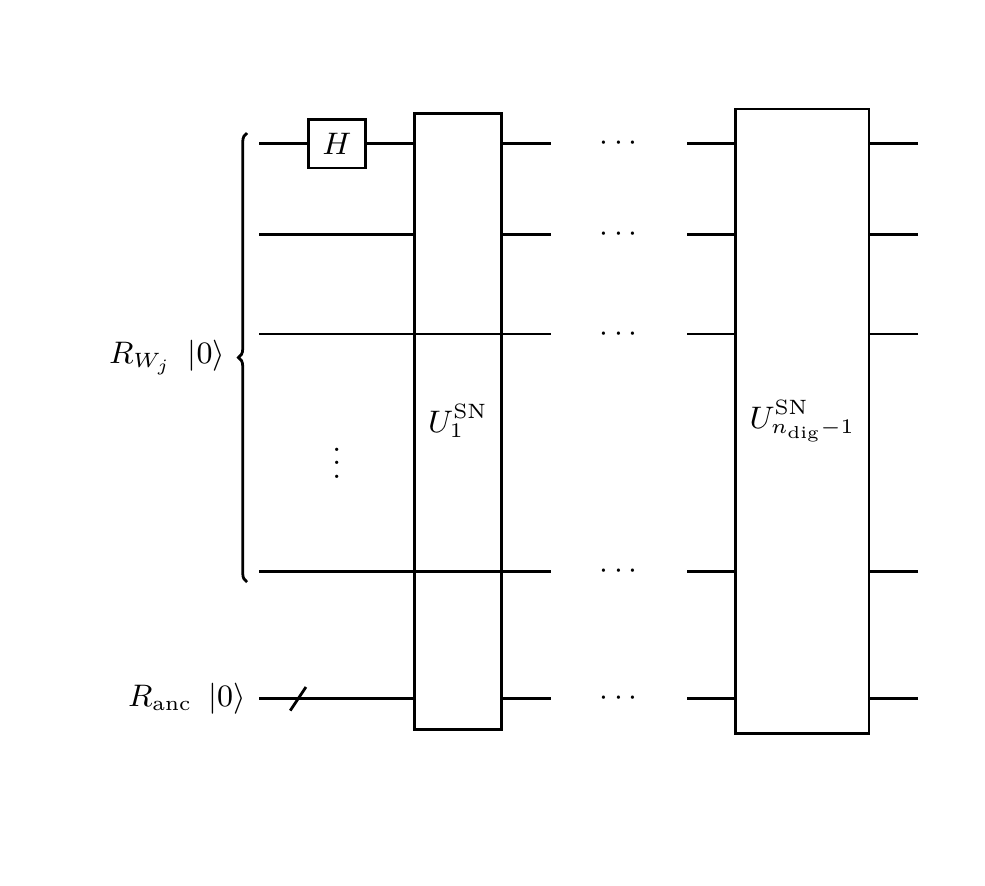}
		\end{center}
		\subcaption{The overview of SN gate.}
	\end{minipage}
	
	\begin{minipage}{1\hsize}
		\begin{center}
			\includegraphics{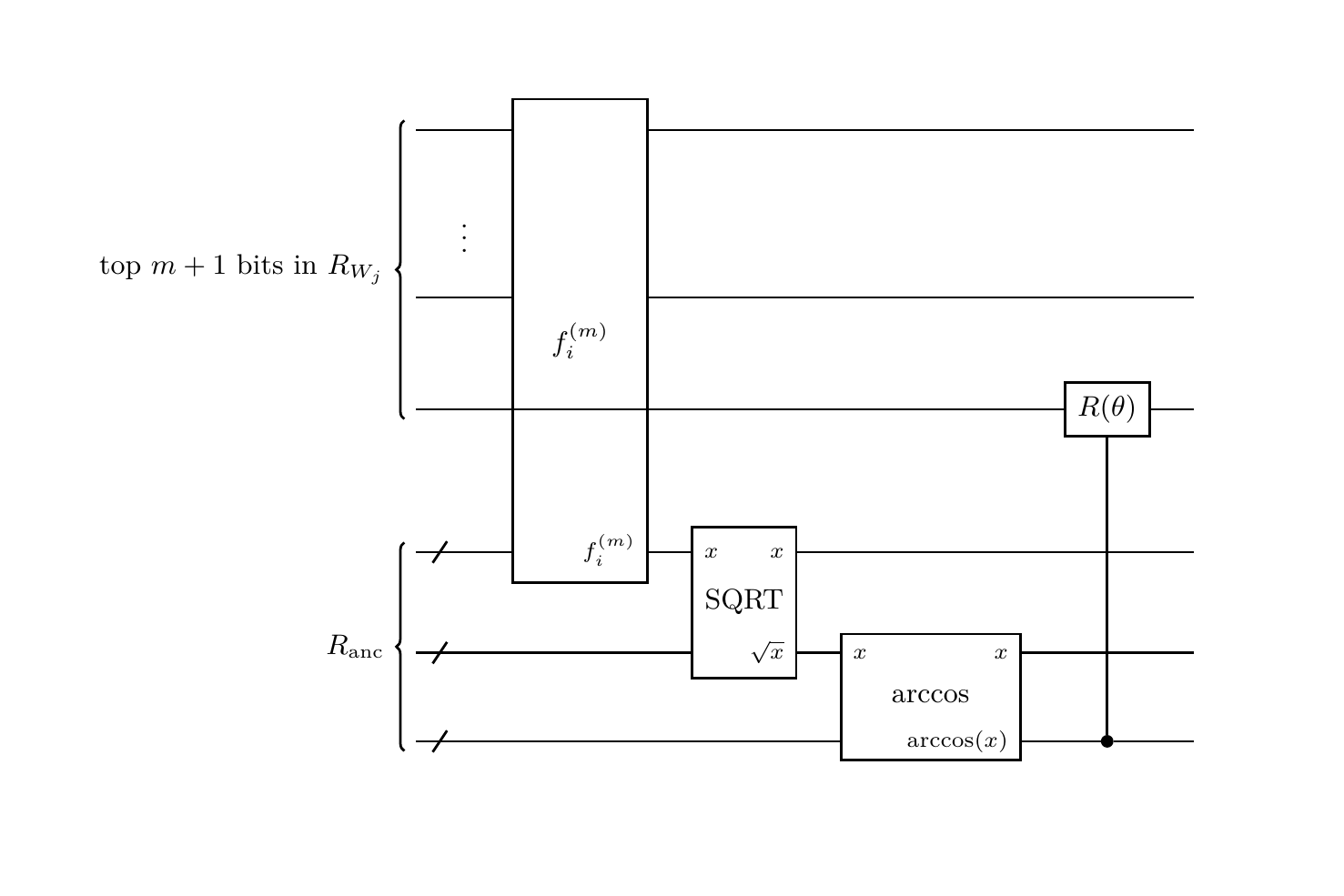}
		\end{center}
		\subcaption{$U^{\rm SN}_m$, the $m$-th step in the SN gate.}
	\end{minipage}
	
	\caption{Implementation of the SN gate.}
	\label{fig:SN}
\end{figure*}

Using this $f^{(m)}_i$ gate, the SN gate is constructed as Figure \ref{fig:SN}.
First, we operate a Hadamard gate to the most significant bit in $R_{W_j}$ to assign probability 1/2 to positive and negative halves of $[x_{{\rm SN},0},x_{{\rm SN},N_{\rm SN}}]$ respectively.
Then, we operate the sequence of the gates $U^{\rm SN}_m$, which corresponds to the $m$-th step of the above recursive calculation.
$U^{\rm SN}_m$ is constructed as a combination of the $f^{(m)}_i$ gate, the gates to calculate square root and arc cosine and the controlled rotation gate $R(\theta)$.

We here comment on implementation of arccos and square root.
The implementation of the inverse trigonometric function based on the piecewise polynomial approximation is proposed in \cite{Haner}.
Although \cite{Haner} considers not arccos but arcsin, these can be easily related as $\arccos(x)=\frac{\pi}{2}-\arcsin(x)$.
We adopt a setting with the polynomial degree 3 and 2 intervals, which leads to accuracy $10^{-5}$\cite{Haner}.
The circuit for square root is given in \cite{MunozCoreas2}.

\section{\label{sec:resources}Estimation of Required Resources}

Then, let us estimate the machine resources required for the implementation in the PRN-on-a-register way and the register-per-RN way.
We consider the two metrics, qubit number and T-count, as many papers do.

\subsection{\label{sec:elemgateres}elementary gates}

\begin{table*}[t]
	\caption{Resources for various elementary gates. We here assume that operands are $n$-bit. We omit subleading terms with respect to $n$.}
	\begin{tabular}{ccccccc} \hline
		\multirow{2}{*}{Gate} & \multicolumn{4}{c}{Qubits} & \multirow{2}{*}{T-count}  &  \multirow{2}{*}{Reference} \\
		 & Total & Operand & Output & Ancilla &  &  \\ \hline \hline
		Adder & $2n$ & $2n$ & 0 (self-update) & 0 & $14n$ & \cite{Thapliyal,Thapliyal2,Thapliyal3} \\  \hline
			Ctrl Adder & $2n$ & $2n$ & 0 (self-update) & 0 & $21n$ & \cite{MunozCoreas,Thapliyal3} \\ \hline
			Modular Adder\footnote{Since we can construct the modular adder using 5 plain adders\cite{Vedral}, we use 5 times the values of the adder for T-count.} & $2n$ &$2n$ & 0 (self-update) & 0 & $70n$ & \cite{Thapliyal,Thapliyal2,Thapliyal3,Vedral} \\  \hline	
				Multiplier & $3n$ & $2n$ & $n$ & 0 & $21n^2$ & \cite{MunozCoreas}\\ \hline
				Divider &$5n$ & $2n$ & $n$ & $2n$ & $35n^2$ & \cite{Thapliyal3} \\ \hline
		Multi Ctrl Toffoli & $2n$ & $n$ & 1 & $n$ & $8n$ & \cite{Selinger,Maslov}\\ \hline
		Square Root\footnote{The circuit given in \cite{MunozCoreas2} takes a $n$-bit input and returns the $n/2$-bit square root and the $n/2$-bit remainder. In order to keep $n$-bit accuracy, we add $n$ 0's to the input and calculate the $n$-bit square root of the virtual $2n$-bit input. We treat the $n$ bits added to the input and the $n$ bits where the remainder is output as ancillas.} & $4n$ & $n$ & $n$ & $2n$ & $14n^2$ & \cite{MunozCoreas2} \\  \hline
		arccos & 105 & ~ & ~ & ~ & $3.4\times 10^4$ \footnote{Since the value shown in \cite{Haner} is Toffoli count, we simply multiply it by 7 for converting it to T-count.}  & \cite{Haner}\\ \hline
		\begin{tabular}{l}controlled rotation \\ (with accuracy of $2^{-n}$) \end{tabular} & \raisebox{0.5em}{2 (control and target)} & ~ & ~ & ~ & \raisebox{0.5em}{$3n$} & \raisebox{0.5em}{\cite{Egger,Kliuchnikov,Amy}} \\ \hline
		\label{tbl:resElem}
	\end{tabular}	
\end{table*}

We first summarize the resources of the elementary gates which are necessary to construct the LV circuit.
We here consider fixed-point arithmetic.
Among various implementations proposed previously, we adopt one for each of the elementary gates and summarize their resources when operands are $n$-bit in Table \ref{tbl:resElem}.
Since we aim to estimate the orders of the above metrics, we take only the leading term with respect to $n$ for required resources.
For example, we approximate $an+b$ as $an$..

We make a comment on multiplication and the division.
For these operation, we use modified versions of circuits proposed in \cite{MunozCoreas} and \cite{Thapliyal3}.
This is because of the following reason.
In order to store the strict value of the product of two $n$-bit numbers, original circuits use $2n$-bits.
This causes a problem in the situation where we have to sequentially perform multiplication as in the LV circuit, since the required qubit number doubles at every multiplication.
Therefore, we have to truncate lower bits of product and keep the digit number constant.
In order to do this, we modify the multiplier in \cite{MunozCoreas}.
We also construct the divider, which is dedicated to drawback of the modified multiplication.
This is why the qubit number for divider in Table \ref{tbl:resElem} is different from that in \cite{MunozCoreas,Thapliyal3}.
We explain the details of the modified multiplier and divider in Appendix.

\subsection{assumptions on registers}

We assume the following points on qubit numbers of various registers, some of which have already been mentioned.
\begin{itemize}
	\item Registers which store numerical numbers, $R_W,R_S,R_{\rm payoff},R_{{\rm LV},j}$ etc., and ancillary registers concerning them have $n_{\rm dig}$ qubits. This is determined according to the required accuracy. We hereafter set $n_{\rm dig}=16$.
	\item $R_{\rm PRN}$ in Figure \ref{fig:JWPW} exceptionally has $n_{\rm PRN}$ qubits, since this is set to so large a value that the PRN sequence has good statistical property, e.g. long period. \cite{PCG} considers the PRN generators with 64 bits and we also use this value for $n_{\rm PRN}$ hereafter. Ancillary registers necessary for calculation of PRN sequence have $n_{\rm PRN}$ qubits too. Even if $n_{\rm PRN}>n_{\rm dig}$, we use only $n_{\rm dig}$ qubits in $R_{\rm PRN}$ for transformation to PSNRN.
	\item $R_{\rm samp}$ has $n_{\rm samp}$ qubits. 
	\item Other registers, e.g. $R_{\rm count}$, have only several qubits and we neglect their contributions to the whole qubit number.
\end{itemize}

\subsection{the PRN-on-a-register way}

Then, let us consider the required resources in the PRN-on-a-register way.

\subsubsection{qubit number}

\begin{table*}[t]
	\caption{ Qubits necessary in each step in the PRN-on-a-register circuit. We neglect registers with only several qubits.}
	\begin{tabular}{cccm{20em}} \hline
		Part & Register & Qubit & Note \\ \hline \hline
		\multirow{4}{*}{Whole} & $R_{\rm samp}$ & $n_{\rm samp}$ & \\ 
	                           & $R_{S}$ & $n_{\rm dig}$ & \\
	                           & $R_{\rm payoff}$ & $n_{\rm dig}$ & \\
	                           & $R_{\rm PRN}$ & $n_{\rm PRN}$ & \\ \hline
	    $J_{\rm PRN}$ & ancilla & $2n_{\rm PRN}$ & To hold intermediate outputs; see (\ref{eq:LCGJump}) \\ \hline
	    \multirow{2}{*}{$\Phi_{\rm SN}^{-1}$} & $R_W$ & $n_{\rm dig}$ & \\  
	                           & \raisebox{1em}{ancilla} & \raisebox{1em}{$6n_{\rm dig}$} & To hold the coefficients of the polynomial and the intermediate outputs; see Figure \ref{fig:InvCDF}\\ \hline
	     \multirow{3}{*}{$V^{(j)}_k$}  & $R_W$ & $n_{\rm dig}$ &  \\
	                            & $R_{S^\prime}$ & $n_{\rm dig}$ & \\  
	                            & \raisebox{1em}{ancilla} & \raisebox{1em}{$4n_{\rm dig}$} & For $x\leftarrow x+(ax+b)y$ and $z\leftarrow \frac{z+x-by}{1+ay}$; see the comment in the body text.\\ \hline
	     $P_{\rm PRN}$ & ancilla & $2n_{\rm PRN}$ & To hold intermediate outputs; see (\ref{eq:ProgTransf}). \\ \hline                       
		\label{tbl:qubit}
	\end{tabular}	
\end{table*}

In Table \ref{tbl:qubit}, we summarize qubits necessary in each step in the circuit.
Registers which hold some values throughout the circuit are as follows: $R_{\rm samp},R_S,R_{\rm payoff}$ ans $R_{\rm PRN}$.
Except these, the following parts in the circuit can consume qubit number most heavily.
\begin{itemize}
	\item $J_{\rm PRN}$ and $P_{\rm PRN}$: $2n_{\rm PRN}$ qubits
	
	\item $\Phi_{\rm SN}^{-1}$:	$7n_{\rm dig}$ qubits
\end{itemize}
Therefore, the total qubit number required in the PRN-on-a-register way is roughly
\begin{equation}
n_{\rm samp} + 2n_{\rm dig} + n_{\rm PRN} + \max\{2n_{\rm PRN},7n_{\rm dig}\}
\end{equation}

Let us comment on some technical points for obtaining Table \ref{tbl:qubit}.
We first make a supplementary explanation on the ancillary qubit number in $V^{(j)}_k$.
There are two parts which require ancillas in  $V^{(j)}_k$.
First, $x\leftarrow x+(ax+b)y$ needs the following ancillas: a $n_{\rm dig}$-bit register to which $1+ay$ is output, a $n_{\rm dig}$-bit register to which the result is temporally output in the self-update multiplication and a $2n_{\rm dig}$-bit register necessary for the inverse division to clear the input $x$.
Second, $z\leftarrow \frac{z+x-by}{1+ay}$ needs the following: a $n_{\rm dig}$-bit register to which $1+ay$ is output and a $2n_{\rm dig}$-bit register necessary for division.
In total, $4n_{\rm dig}$ bits are sufficient\footnote{Strictly speaking, comparisons between $R_S$ or $R_{S^\prime}$ and $s_{j,k}$'s require loading $s_{j,k}$'s into some register. This does not require another register, since at least one of ancillary registers used in $x\leftarrow x+(ax+b)y$ and $z\leftarrow \frac{z+x-by}{1+ay}$ is empty at loading. }.

We also comment on the ancilla number in $\Phi_{\rm SN}^{-1}$.
As we can see from Figure \ref{fig:InvCDF}, we need four registers to which coefficients are loaded and two registers for intermediate outputs.
Therefore, $6n_{\rm dig}$ ancillas are necessary\footnote{Although we also need a register to which $x^{\rm ICDF}_{m}$'s are loaded at comparisons between them and $R_{\rm PRN}$, we can use $R_W$ or intermediate output registers, which are empty at comparisons.}

\subsubsection{T-count}

Since we are interested in only the leading contribution, we focus on multiplications, divisions and repeated additions.
Besides, we do not consider the T-count of $J_{W}$, which is used only once.
For the parts in $U_j$, which is used repeatedly, we specify T-counts as follows:

\begin{enumerate}
	\item $V^{(j)}_k$ \\
	One $V^{(j)}_k$ includes the following parts:
	\begin{itemize}
		\item $x\leftarrow x+(ax+b)y$ \\
		As we can see in (\ref{eq:axby}), this includes one multiplication and one division, which come from one self-update multiplication, and $3n_{\rm dig}$ controlled additions, which comes from two controlled multiplications by constant and one inverse.
		In total, the T-count is $119n_{\rm dig}^2$.
		
		\item $z\leftarrow \frac{z+x-by}{1+ay}$ \\
		As we can see in (\ref{eq:divInVjk}), this includes one division and $2n_{\rm dig}$ additions, which comes from two multiplications by constant.
		In total, the T-count is $63n_{\rm dig}^2$.
		
		\item Uncomputation of $z\leftarrow \frac{z+x-by}{1+ay}$ \\
		Similar to the above.
	\end{itemize}
    Therefore, the total T-count in one $V^{(j)}_k$ is $245n_{\rm dig}^2$.
    Since $V^{(j)}_k$ is used $n_S+1$ times, the total T-count in them is $245n_{\rm dig}^2n_S$ (only the leading term).

	\item $P_{\rm PRN}$ \\
	This includes two modular multiplications by constant, which comes from one self-update modular multiplication.
	These are decomposed into $2n_{\rm PRN}$ modular additions.
	So the T-count is roughly $140n_{\rm PRN}^2$.
	
	\item $\Phi_{\rm SN}^{-1}$ and its inverse \\
	Each of them includes $2(n_{\rm ICDF} + 1)$ additions ($n_{\rm ICDF} + 1$ comparisons) and five multiplications.
	So the T-count for each is roughly $105n_{\rm dig}^2 + 28n_{\rm dig}n_{\rm ICDF}$.

\end{enumerate}
Summing up these and considering $U_j$ is used in $n_t$ times, the T-count in the whole circuit is roughly
\begin{equation}
(245n_{\rm dig}^2n_S+140n_{\rm PRN}^2+210n_{\rm dig}^2 + 56n_{\rm dig}n_{\rm ICDF})n_t.
\end{equation}

\subsection{the register-per-RN way}

Next, we consider the required resources in the register-per-RN way.

\subsubsection{qubit number}

\begin{table*}[t]
	\caption{ Qubits added in each time step in the register-per-RN circuit. We neglect registers with only several qubits. We only take the leading contributions.}
	\begin{tabular}{lm{16em}ll} \hline
		\multicolumn{2}{l}{Register} & Qubit & Note \\ \hline \hline
		\multicolumn{2}{l}{$R_{S_i}$} & $n_{\rm dig}$ & \\ \hline
		\multicolumn{2}{l}{$R_{W_i}$} & $n_{\rm dig}$ & \\ \hline
		\multicolumn{2}{l}{$R_{{\rm payoff},i}$} & $n_{\rm dig}$ & \\  \hline
		\multicolumn{2}{l}{$R_{{\rm LV},i}$} & $2n_{\rm dig}$ & \\ \hline
		\multicolumn{2}{l}{ancilla used for $x\leftarrow x+(ax+b)y$ in $U_{t_i}$} & $n_{\rm dig}$ & See (\ref{eq:axbyRN})\\ \hline
		\multirow{3}{*}{ancilla for $U^{\rm SN}_m,m=1,...,n_{\rm dig}-1$} &  output $f^{(m)}_i$ & $n_{\rm dig}^2$ & $n_{\rm dig}$ for one $U^{\rm SN}_m$  \\
		& ancilla for SQRT & $2n_{\rm dig}^2$ & $2n_{\rm dig}$ for one $U^{\rm SN}_m$ \\
		& \begin{tabular}{l}qubits used for arccos \\ (input, output and intermediate output)\end{tabular} & $105n_{\rm dig}$ & 105 for one $U^{\rm SN}_m$ \\  \hline                       
		\label{tbl:qubitRN}
	\end{tabular}	
\end{table*}

In the register-per-RN way, registers shown in Table \ref{tbl:qubitRN} are added per time step.
Note that we do not uncompute ancillas.
Summing up all registers, the qubit number necessary for one time step is roughly $3n_{\rm dig}^2+111n_{\rm dig}$, and for the entire circuit it is
\begin{equation}
(3n_{\rm dig}^2+111n_{\rm dig})n_t.
\end{equation}
Note that the dominant part comes from the iterative calculation in the SN gates, which prepare superpositions of the values of the SNRNs.

\subsubsection{T-count}

Again, we focus on operations with large T-count.
For each part in the circuit, we estimate the T-count as follows:

\begin{enumerate}
	\item SN gate \\
	The $m$-th iteration $U^{\rm SN}_m$ in the SN gate includes the following parts:
	\begin{itemize}
		\item square root, arccos, controlled rotation \\
		T-counts are $14n_{\rm dig}$, $3.4\times 10^4$ and $3n_{\rm dig}$, respectively.
		
		\item $f^{(m)}_i$ \\
		For $2\le m\le 6$, we use $2^m$ $m$-controlled Toffoli gates to check the value on $R_{W_i}$ and load $f^{(m)}_i$ which corresponds to the value.
		T-count for this is $2^m(8m-9)$\footnote{Here, we use $8m-9$, the accurate value of T-count of the $m$-controlled Toffoli gates\cite{Selinger,Maslov}, since the approximation as $8m$ is too crude for small $m$.}.
		Summing this for $m=2,...,6$ leads to about 4000.
		Since this is much smaller than T-count for arccos in one iteration, we neglect this.
		For $m\ge 7$, we do multiplication between a $m$-bit variable and a $n_{\rm dig}$-bit constant, which is decomposed $n_{\rm dig}$ additions of $m$-bit.
		Then, T-count is $14mn_{\rm dig}$.
		
	\end{itemize}
	Summing up these and taking only dominant contributions, one SN gate has T-count of $(7n_{\rm dig}^2 + 3.4\times 10^4)n_{\rm dig}$ roughly.

	\item $U_j$ \\
	This includes $2n_S$ additions ($n_S$ comparisons) and three multiplications.
	So one $U_j$ gates has T-count of $63n_{\rm dig}^2+28n_Sn_{\rm dig}$ roughly.
	
\end{enumerate}
In total, we can estimate T-count of the entire circuit in the register-per-RN way as
\begin{equation}
(7n_{\rm dig}^2 + 63n_{\rm dig} + 28n_S+3.4\times 10^4)n_{\rm dig}n_t.
\end{equation}

\subsection{comparison between two ways}

\begin{table*}[t]
	\caption{ Qubit number and T-count required in the PRN-on-a-register and register-per-RN ways.}
	\begin{tabular}{cm{2em}cm{2em}c} \hline
		 & \quad & PRN-on-a-register & \quad & register-per-RN \\ \hline \hline
		qubit & \quad & $n_{\rm samp} + 2n_{\rm dig} + n_{\rm PRN} + \max\{2n_{\rm PRN},7n_{\rm dig}\}$ & \quad & $(3n_{\rm dig}^2+111n_{\rm dig})n_t$ \\ \hline
		T-count & \quad & $(245n_{\rm dig}^2n_S+140n_{\rm PRN}^2+210n_{\rm dig}^2 + 56n_{\rm dig}n_{\rm ICDF})n_t$ & \quad & $(7n_{\rm dig}^2 + 63n_{\rm dig} + 28n_S+3.4\times 10^4)n_{\rm dig}n_t$ \\ \hline                    
		\label{tbl:ressum}
	\end{tabular}	
\end{table*}

\begin{table*}[t]
	\caption{ Required resources to implement the PRN-on-a-register and register-per-RN ways in the valid case (\ref{eq:nSpec}). The following values are obtained from Table \ref{tbl:ressum}.}
	\begin{tabular}{ccc} \hline
	 & PRN-on-a-register & register-per-RN \\ \hline
		qubit  & $2.4\times 10^2$ & $9.2\times 10^5$ \\
		T-count & $3.7\times 10^8$ & $2.1\times10^8$ \\ \hline                       
		\label{tbl:ressumspec}
	\end{tabular}	
\end{table*}

We then compare resources necessary in two ways in Table \ref{tbl:ressum}.
Naturally, qubit number is independent from $n_t$ in the PRN-on-a-register way but proportional to $n_t$ in the register-per-RN way and T-count is proportional to $n_t$ in the both ways.
If we take a setting, which is typically necessary for practical use in derivative pricing\footnote{$n_{\rm samp}=16$ corresponds to 65536 sample paths.} :
\begin{eqnarray}
n_{\rm samp} &=& 16, \nonumber \\
n_{\rm dig} & = & 16, \nonumber \\
n_{\rm PRN} & = & 64, \nonumber \\
n_{\rm ICDF} & = & 109, \nonumber \\
n_t & = & 360, \nonumber \\
n_S & = & 5 \label{eq:nSpec}
\end{eqnarray}
the values in Table \ref{tbl:ressum} becomes as Table \ref{tbl:ressumspec}.
The total T-count is of same order of magnitude in the both way but larger for the PRN-on-a-register way by a factor 2 roughly.
We here comment on the parts which consume T-count most heavily in each way.
In the PRN-on-a-register way, there are two parts which contribute to T-count equally and dominantly.
The first is the update of the asset price in $V^{(j)}_k$.
Note that additional operations for reduction of qubits, such as inverse division in self-update multiplication and drawing back the asset price to clear $R_g$, increase T-count compared with the register-per-RN way.
The second is modular multiplications in update of the PRN sequence.
Since the PRN generator requires the large bit number, say $n_{\rm PRN}=64$, in order to it keep good statistical properties such as long period, the T-count of operations for the PRN becomes large.
On the other hand, in the register-per-RN way the dominant contribution to T-count comes from arccos's in preparing SNRNs.
Because not only an arccos itself is T-count consuming but also it is used in each iteration in the SN gate, the total T-count piles up.

\section{\label{sec:con}Summary}

In this paper, we considered how to implement time evolution of the asset price in the LV model on quantum	computers.
Similar to other problems in finance, derivative pricing by Monte Carlo simulation requires a large number of random numbers, which is proportional to $n_t$, the number of time steps for asset price evolution, and this may cause difficulty in implementation.
We then considered two ways of implementation: the PRN-on-a-register way and the register-per-RN way.
In the former we sequentially generate pseudo random numbers on a register and use them to evolve the asset price. 
In the latter, standard normal random numbers necessary to time evolution are created as superpositions on separate registers.
For both ways, we present the concrete quantum circuits in detail.
For not only random number generation but also other aspects, we try to save qubit numbers permitting some additional procedures in the PRN-on-a-register way and do opposite in the register-per-RN way.
We then give estimations of qubit number and T-count required in each way.
In the PRN-on-a-register way, qubit number is kept constant against $n_t$.
On the other hand, in the register-per-RN way qubit number is proportional to $n_t$.
Each way has T-count consuming parts and the total T-counts for both ways are of same order of magnitude, expect the PRN-on-a-register way has the larger T-count by a factor about 2, in some specific setting.

Note that analyses of resources required for implementation of the LV model in this paper strongly depend on designs of elementary circuits for arithmetic.
For example, in the register-per-RN way the dominant contribution to T-count comes from arccos's in preparing SNRNs.
If more efficient circuits are proposed and we replace the current choice with them, required resources may change.

Finally, we would like to notice that this study is not enough for application of quantum algorithm for Monte Carlo simulation to pricing in the LV model.
Although we assumed that the LV function is given, in practice we have to calibrate the LV so that the model prices of European options fit to the market prices.
Besides, we have not considered how to evaluate terms in exotic derivatives, for example, early exercise.
In future works, we will consider such things and aim to present how to apply quantum computers in the whole process of exotic derivative pricing.

\appendix

\section{Sufficient condition on $\sigma(t,S)$ in the PRN-on-a-resigter way}

We here show that $\sigma(t,S)$ which is continuous with respect to $S$ and takes the form of (\ref{eq:LV}) and sufficiently small $\Delta t_j$ lead to one-to-one correspondence between $S^{(i)}_{t_j}$ and $S^{(i)}_{t_{j+1}}$.
We see $S^{(i)}_{t_{j+1}}$ as a function of $S^{(i)}_{t_{j}}$ and define a function $f$ by
\begin{equation}
f(S) =  S + \sigma(t_j,S)\sqrt{\Delta t_j}w_j \label{eq:fS}
\end{equation}
for fixed $w_i$ so that $S^{(i)}_{t_{j+1}}=f(S^{(i)}_{t_j})$ holds.
Except for the grid points $S=s_{j,0}...s_{j,n_S}$, $f(S)$ is differentiable and
\begin{equation}
f^\prime(S)=1+a_{j,k}\sqrt{\Delta t_j} w_j; \ {\rm for} \ s_{j,k-1}<S<s_{j,k},k=0,...,n_S+1.
\end{equation}
Since $w_j$ is bounded in numerical computation, $f^\prime(S)$ is always positive expect the grid points for sufficiently small $\Delta t_j$.
Besides, if $\sigma(t_j,S)$ is continuous with respect to $S$ also at the grid points, so is $f(S)$.
Combining these, we find that $f(S)$ is strictly increasing for small $\Delta t_j$.
Thus, the correspondence between $S^{(i)}_{t_j}$ and $S^{(i)}_{t_{j+1}}$ is one-to-one.

Small $\Delta t_j$ is required from another perspective, too.
The original dynamics of the asset price is given as the continuous-time evolution (\ref{eq:SDE}) and (\ref{eq:EM}) is the discretized approximation of it.
Therefore, in order for this approximation to be accurate, the increment of asset price should be sufficiently smaller than the asset price itself: $\left|\sigma(t_j,S^{(i)}_{t_j})\sqrt{\Delta t_j}w_j\right| \ll \left|S^{(i)}_{t_j}\right|$.
If this condition is satisfied, $\left|a_{j,k}\sqrt{\Delta t_j} w_j \right| \ll 1$, so $f^\prime(S)>0$ is met. 

\section{Truncated multiplier and divider}

We here describe the modified version of multiplier and divider.
We assume that we consider the fixed-point arithmetic with $n_{\rm int}$ bits in the integer part and $n_{\rm frac}$ bits in the fractional part, $n=n_{\rm int}+n_{\rm frac}$ bits in total.
We hereafter call such numbers $(n_{\rm int},n_{\rm frac})$-bit numbers.
We want to keep this digit setting before and after multiplication.
In order to do this, we adopt the following policy.

\begin{figure*}[t]
	\begin{center}
		\includegraphics[width=0.9\textwidth]{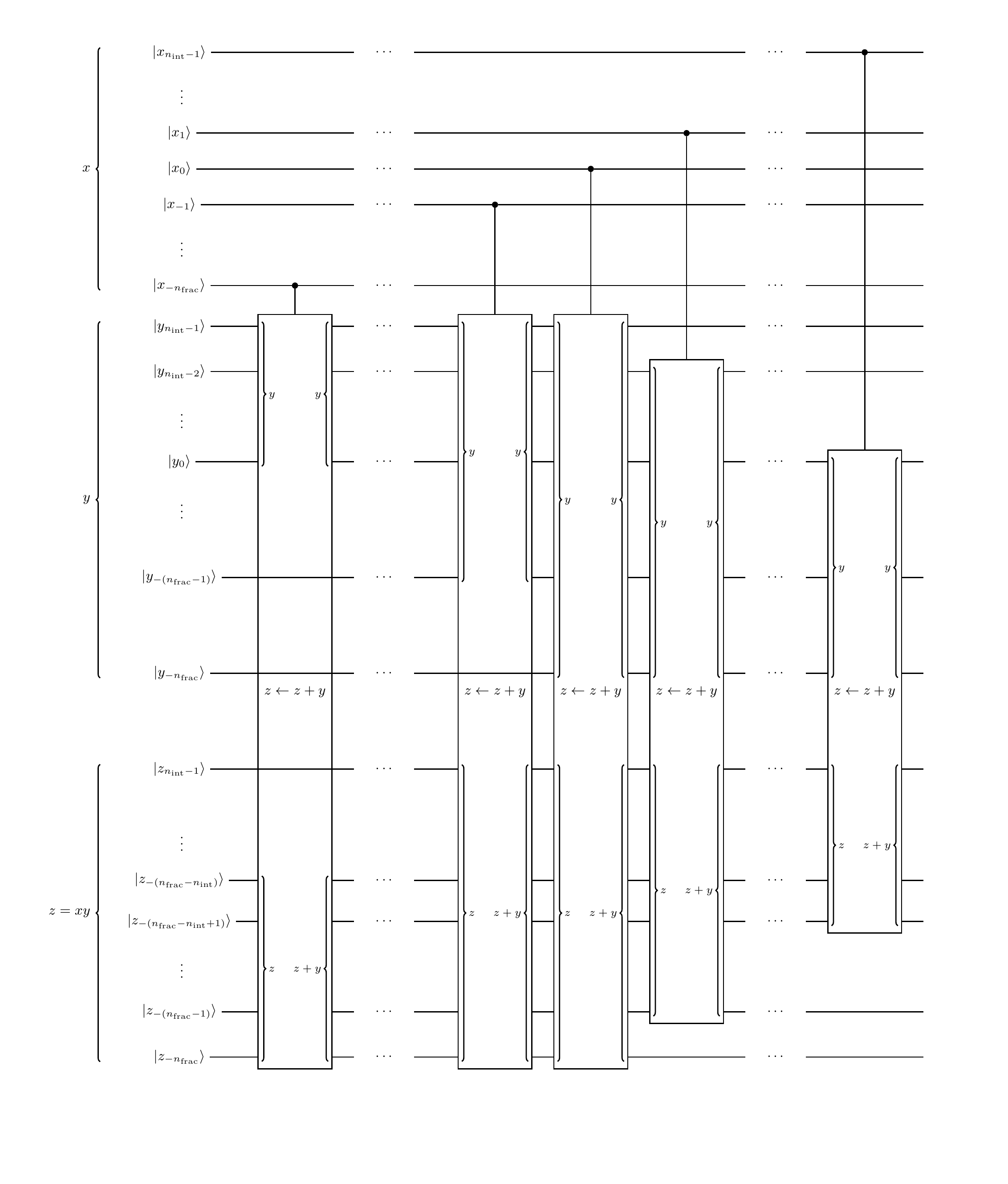}
	\end{center}
	\caption{The circuit to perform truncated multiplication.}
	\label{fig:xytrunc}
\end{figure*}

\begin{itemize}
	\item We simply truncate the digits lower than the $n_{\rm frac}$-th fractional digit in the product. This might cause numerical errors around and the $n_{\rm frac}$-th fractional digit and such a tiny error might accumulate, but we simply neglect this concern.
	\item We assume the overflow from the $n_{\rm int}$-bit integer part never occurs.
\end{itemize}
We then approximate the product as follows.
Writing a number $x$ in binary representation as $x_{n_{\rm int}-1}x_{n_{\rm int}-2}...x_0.x_{-1}...x_{-n_{\rm frac}}$, where $x_i$ is the $i$-th integer digit of $x$ and $x_{-j}$ is the $j$-th fractional digit of $x$, we do
\begin{equation}
xy = \sum_{i=-n_{\rm frac}}^{n_{\rm int}-1} {x_i2^iy}
 \approx  f^{\rm mul}_{n_{\rm frac},n_{\rm int},y}(x)
 :=  \sum_{i=-n_{\rm frac}}^{n_{\rm int}-1} {x_i2^i\tilde{y}_i},
\end{equation}
where
\begin{equation}
\tilde{y}_i=
\begin{cases}
y_{n_{\rm int}-1}...y_0.y_{-1}...y_{-(n_{\rm frac}-i)} \  ; \ {\rm for} \ i< 0 \\
y \ ;  \ {\rm for} \ i\ge 0
\end{cases}.
\end{equation}
This truncated multiplication is implemented as a circuit in Figure \ref{fig:xytrunc}.
Note that the circuit in Figure \ref{fig:xytrunc} actually calculates not $f^{\rm mul}_{n_{\rm frac},n_{\rm int},y}(x)$ but 
\begin{equation}
\sum_{i=0}^{n_{\rm int}-1} {x_i2^i(y_{n_{\rm int}-1-i}...y_0.y_{-1}...y_{-n_{\rm frac}})} + \sum_{j=1}^{n_{\rm frac}} {x_{-j}2^{-j}(y_{n_{\rm int}-1}...y_0.y_{-1}...y_{-(n_{\rm frac}-j)})}.
\end{equation}
This is equal to $f^{\rm mul}_{n_{\rm frac},n_{\rm int},y}(x)$ if our assumption that the overflow from the $n$-bit integer part never occurs is satisfied. 

\begin{figure*}[t]
	
		\begin{minipage}{1\hsize}
	\begin{center}
	\includegraphics[width=0.85\textwidth]{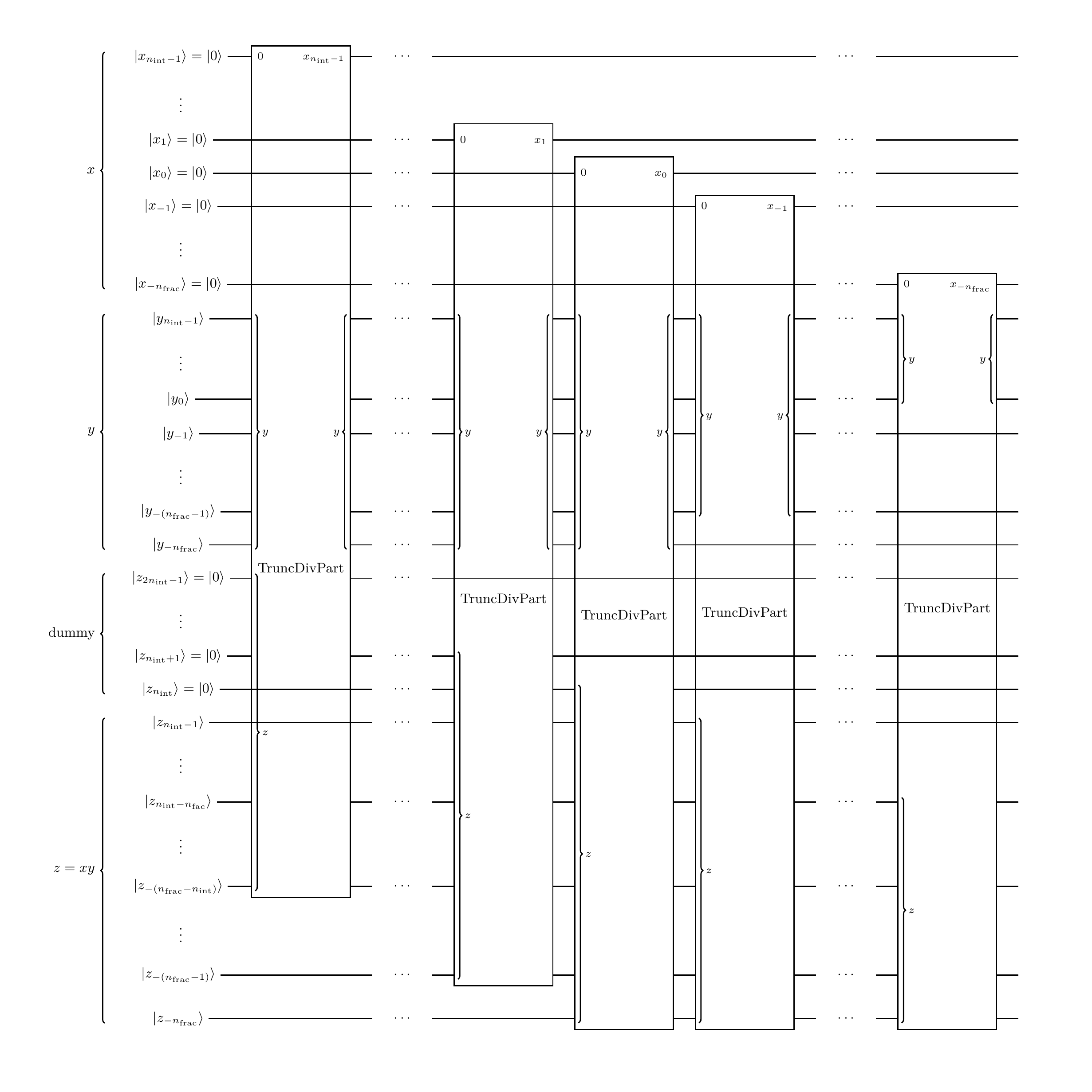}
\end{center}
		\subcaption{The overview.}
	\end{minipage}
	
			\begin{center}
	\begin{minipage}{1\hsize}

			\includegraphics[width=1\textwidth]{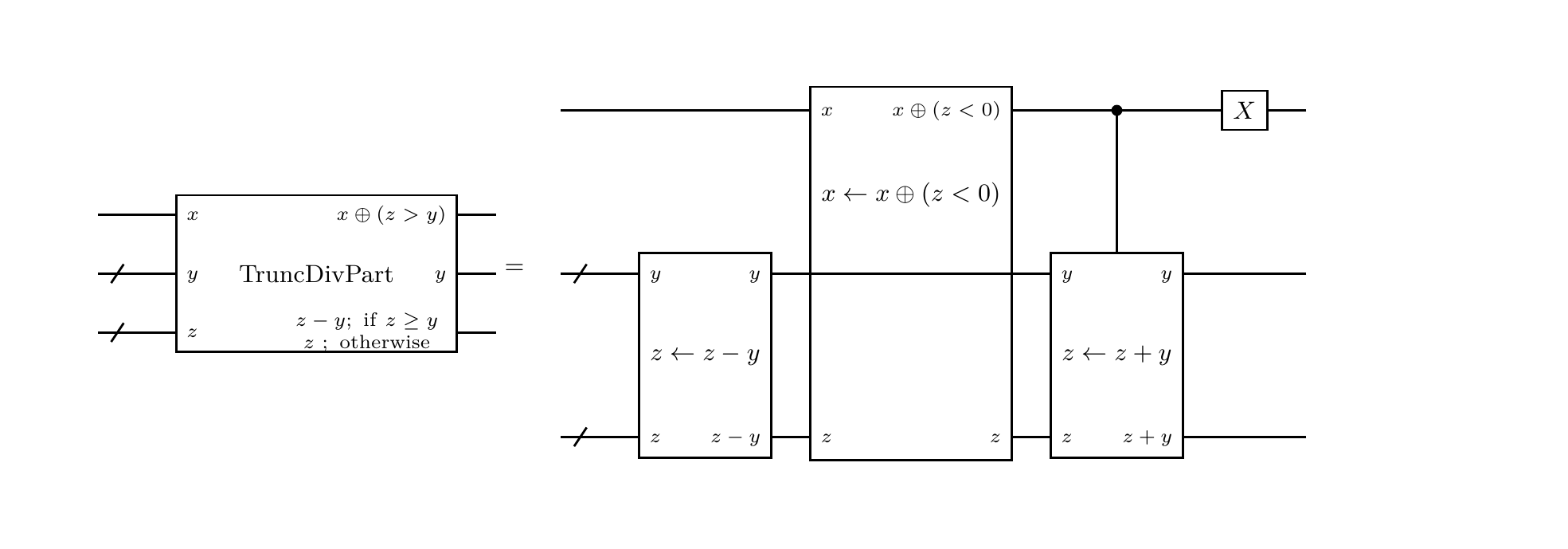}
		\subcaption{The TruncDivPart gate. Note that in the 2's complement method we can check whether a number is negative or not by seeing the most significant bit so we do not have to use an adder as a comparator.}
		\end{minipage}
			\end{center}

\caption{The circuit for the truncated division.}
	\label{fig:truncdiv}
\end{figure*}

We define the truncated division as the inverse of the truncated multiplication: $z/y\approx f^{\rm div}_{n_{\rm frac},n_{\rm int},y}(z):=\left(f^{\rm mul}_{n_{\rm frac},n_{\rm int},y}\right)^{-1}(z)$.
Given two $(n_{\rm int},n_{\rm frac})$-bit numbers $y,z$ which satisfies $z=f^{\rm mul}_{n_{\rm frac},n_{\rm int},y}(x)$, we can find the $(n_{\rm int},n_{\rm frac})$-bit number $x$ as follows:
\begin{enumerate}
	\item Set $i=n_{\rm int}-1$, $d=z$ and $x=0$.
	\item Update $d$ with $d-2^{i}\tilde{y}_i$
	\item If $d<0$, update $d$ with $d+2^{i}\tilde{y}_i$ ($d$ returns to the value before step 2) and set $x_i=0$. If $d\ge 0$, set $x_i=1$.
	\item Decrement $i$ by 1.
	\item Repeat step 2-4 until $i$ becomes $-n_{\rm frac}-1$ at step 4.
	\item Output $x$.
\end{enumerate}
Note that $2^{i}\tilde{y}_i > \sum_{j=-n_{\rm frac}}^{i-1}2^{j}\tilde{y}_j$.
This ensures that sequential subtractions by $2^{i}\tilde{y}_i$ and checking whether the difference is positive or negative lead to determining each digit of $x$.
The above procedure is implemented as the circuit in Figure \ref{fig:truncdiv}.
Note that we add dummy qubits which correspond to the $2n_{\rm int}-1$-th to $n_{\rm int}$-th integer digits of $z$.
Also note that this circuit transforms the dividend register from $\ket{z}$ to $\ket{0}$.
If we want to reserve $\ket{z}$, we can copy the state to another ancillary register by CNOT gates and use the copy as the dividend state.
That is, we can do
\begin{equation}
\ket{z}\ket{0}\ket{y}\ket{0} \rightarrow\ket{z}\ket{z}\ket{y}\ket{0}\rightarrow\ket{z}\ket{0}\ket{y}\ket{x}.
\end{equation} 

Despite the trick to truncate the digits, the structures of the circuits for truncated multiplication and division are similar to that in \cite{MunozCoreas} and the restoring division circuit in \cite{Thapliyal3} respectively.
We therefore use the qubit number and T-count of the circuits in \cite{MunozCoreas,Thapliyal3} as those of our truncated arithmetic circuits.
The exception is the qubit number of divider, for which we use $5n$.
This is larger than the value in \cite{Thapliyal3} by $2n$, reflecting the addition of dummy qubits and the register to which the dividend is copied\footnote{Actually, added qubits are not $2n$ but $n_{\rm int}+n$, but we consider that $2n$ qubits are added for simplicity and conservativeness. }.

\end{document}